\documentclass[12pt,english,longbibliography,nofootinbib,superscriptaddress,12pt,sort&compress,showkeys]{revtex4-1}
\usepackage{ae,aecompl}
\usepackage[T1]{fontenc}
\usepackage[latin9]{inputenc}
\usepackage[active]{srcltx}
\usepackage{amsmath}
\usepackage{graphicx}
\usepackage{grffile}

\makeatletter

\providecommand{\tabularnewline}{\\}

\usepackage{aecompl}\usepackage{epstopdf}

\usepackage{babel}

\makeatother

\usepackage{babel}
\begin{document}
\title{Stochastic model and kinetic Monte Carlo simulation of solute interactions
with stationary and moving grain boundaries. II. Application to two-dimensional
systems}
\author{Y. Mishin}
\address{\noindent Department of Physics and Astronomy, MSN 3F3, George Mason
University, Fairfax, Virginia 22030, USA}
\begin{abstract}
\noindent In Part I of this work, we proposed a stochastic model describing
solute interactions with stationary and moving grain boundaries (GBs)
and applied it to planar GBs in 1D systems. The model reproduces nonlinear
GB dynamics, solute saturation in the segregation atmosphere, and
all basic features of the solute drag effect. Part II of this work
extends the model to 2D GBs represented by solid-on-solid interfaces.
The model predicts a GB roughening transition in stationary GBs and
reversible dynamic roughening in moving GBs. The impacts of the GB
roughening on GB migration mechanisms, GB mobility, and the solute
drag are studied in detail. The threshold effect in GB dynamics is
explained by the dynamic roughening transition, which is amplified
in the presence of solute segregation. The simulation results are
compared with the classical models by Cahn and Lücke-Stüwe and previous
computer simulations.
\end{abstract}
\keywords{Grain boundary, solute drag, kinetic Monte Carlo, grain boundary roughening
transition}
\maketitle

\section{Introduction}

In many alloys, the chemical components strongly interact with grain
boundaries (GBs), reducing their mobility and, in some cases, pinning
the GBs in place. The most common mechanism of the solute-induced
retardation of GB motion is the solute drag effect, in which the GB
moves slower when it carries a solute segregation atmosphere \citep{Balluffi95}.
As a result, a larger driving force must be applied to sustain the
GB motion compared with the force required to move the GB with the
same velocity in the pure solvent. The difference between the two
forces is called the solute drag force, and its strength is controlled
by competition between GB migration and solute diffusion. If the solute
diffusivity is high, a heavy atmosphere is dragged by the moving GB,
drastically reducing its mobility. If the solute diffusivity is low,
the GB can break away from the atmosphere and move faster.

The classical solute drag models by Cahn \citep{Cahn-1962} and Lücke
et al.~\citep{Lucke-Stuwe-1963,Lucke:1971aa}, and more recent computer
simulations \citep{Mendelev:2001aa,Mendelev02a,Wang03,Gronhagen:2007aa,Abdeljawad:2017aa,Kim:2008aa,Li:2009aa,Greenwood:2012aa,Shahandeh:2012aa,Wicaksono:2013aa,Sun:2014aa,Rahman:2016aa,Mishin:2019aa,Koju:2020aa,Alkayyali:2021uo,Koju:2021aa,Suhane:2022aa},
predict a highly nonlinear relation between the GB velocity $v$ and
the solute drag force $F_{d}$. In particular, they predict a maximum
of $F_{d}$ at a critical velocity $v_{*}$ separating two kinetic
regimes: the segregation drag at $v<v_{*}$ and a breakaway from the
atmosphere at $v>v_{*}$. Several open questions remain in this field.
For example, Cahn \citep{Cahn-1962} predicted a morphological instability
of the moving GB in the breakaway regime, which was not observed in
simulations. It is also known that GBs can undergo a roughening transition
at high temperatures \citep{Swendsen:1977,Rottman86a,Olmsted07a,Baruffi:2022aa}.
There is evidence that GB roughness increases GB mobility. However,
it is less clear how the GB motion affects the roughness. Furthermore,
the impact of the roughening transition on the solute drag effect
remains unexplored.

In Part I of this work \citep{Mishin_2023_RW_part_I}, we proposed
a simple stochastic model describing solute interactions with stationary
and moving GBs. The model is solved by kinetic Monte-Carlo (KMC) simulations
with time-dependent transition barriers. The time dependence captures
the increase in the GB displacement barriers when the solute atoms
diffuse towards the GB to form a segregation atmosphere. The increasing
barriers reduce the GB mobility in a process that we call GB \emph{pinning}
\citep{Mishin_2023_RW_part_I}. The model was applied to a planar
GB driven by an external force \citep{Mishin_2023_RW_part_I}. It
was shown that the model reproduces all basic features of the solute
drag effect, including the maximum of the drag force at a critical
velocity. By contrast to the classical models \citep{Cahn-1962,Lucke-Stuwe-1963,Lucke:1971aa},
which also assume planar GB geometry, the present model describes
nonlinear GB dynamics and the solute saturation in the segregation
atmosphere. While the classical models predict that the maximum drag
force must be independent of the solute diffusivity, the simulations
have shown a significant increase in the maximum drag force with increasing
diffusivity. This increase should be expected: when the solute diffusivity
is fast, the segregation atmosphere can follow the moving GB up to
higher velocities, extending the solute drag branch of the force-velocity
relation towards larger drag forces.

In the present paper (Part II), we extend the model to 2D systems.
This will allow us to study the GB shape fluctuations in the form
of either kink pairs or capillary waves. In section \ref{sec:2D-model},
we formulate the 2D version of the model representing the GB as a
solid-on-solid interface with an adjustable interface energy. The
model reproduces a roughening transition in both stationary and moving
boundaries. This allows us to study the dynamic roughening effect
and its impact on GB migration mechanisms and the solute drag process. 

\section{model formulation and simulation method\label{sec:2D-model}}

In the 2D version of the model, the GB is a 1D object (curve) separating
two 2D grains. The model is illustrated in Fig.~\ref{fig:2D-interface}(a).
The GB is composed of $N$ straight segments connecting the nodes
of an imaginary $a\times a$ grid. The nodes are at $x_{i}=ia$, $z_{i}=ja$,
where $i=0,1,...,N-1$ and $j$ are integers. The periodic boundary
condition $z_{N}=z_{0}$ is imposed. The nodes can be interpreted
as structural units of the GB. Each GB segment $[i,i+1]$ is assigned
the excess energy
\begin{equation}
\varepsilon_{i}=\gamma a\left[\sqrt{a^{2}+(z_{i+1}-z_{i})^{2}}-a\right],\label{eq:sos_1}
\end{equation}
where $\gamma$ is the GB energy per unit area assumed to be the same
for all segments. For a planar interface all $\varepsilon_{i}=0$. 

Each GB node $i$ is acted upon by two forces: (1) external force
$F$ applied parallel to the $z$-axis, and (2) local interface tension
$\varepsilon_{i-1}+\varepsilon_{i}$. At a finite temperature, each
node executes a driven random walk along the $z$-axis. 

The model falls in the category of solid-on-solid (SOS) models \citep{Swendsen:1977,Lapujoulade:1994uw,Hasenbusch:1996wn,LandauBinder09},
which were originally developed for surface roughening and crystal
growth from a vapor phase. SOS models have several versions, depending
on the algorithm for computing the excess energy. The best-known of
them are the discrete Gaussian SOS (DGSOS) model with
\begin{equation}
\varepsilon_{i}=\gamma(z_{i+1}-z_{i})^{2},\label{eq:sos_2}
\end{equation}
and the absolute SOS (ASOS) model with 
\begin{equation}
\varepsilon_{i}=\gamma a\left|z_{i+1}-z_{i}\right|.\label{eq:sos_3}
\end{equation}
A 3D DGSOS model was recently used to simulate solute drag by moving
GBs \citep{Wicaksono:2013aa}. There is no compelling physical reason
to prefer one SOS version over another. The ansatz in Eq.(\ref{eq:sos_1})
interpolates between the DGSOS and ASOS versions. It converges to
the ASOS version when $\left|z_{i+1}-z_{i}\right|\gg a$ but regularizes
the discontinuity of the energy derivative with respect to the inclination
angle at $z_{i}=z_{i+1}$.

To describe the GB dynamics, we adopt the harmonic transition state
theory (TST) \citep{Vineyard:1957vo}, by which the forward ($+$)
and backward ($-$) transition (jump) rates of any GB node $i$ are
$\omega_{i}^{\pm}=\nu_{0}P_{i}^{\pm}$, where $\nu_{0}$ is the attempt
frequency assumed to be constant,
\begin{equation}
P_{i}^{\pm}=\exp\left(-\dfrac{E_{i}^{\pm}}{k_{B}T}\right)\label{eq:1.01-1}
\end{equation}
are the jump probabilities, and $k_{B}$ is Boltzmann's constant.
The jump barriers $E_{i}^{\pm}$ are given by
\begin{equation}
E_{i}^{\pm}=\begin{cases}
E_{ti}\exp\left(\dfrac{u_{i}^{\pm}}{2E_{ti}}\right), & u_{i}^{\pm}\leq0,\\
u_{i}^{\pm}+E_{ti}\exp\left(-\dfrac{u_{i}^{\pm}}{2E_{ti}}\right), & u_{i}^{\pm}>0,
\end{cases}\label{eq:1.02-1}
\end{equation}
where $u_{i}^{\pm}$ is the total energy change due to the jump. This
energy change includes the work $\mp Fa$ against the external force
$F$ and the energy changes, $\Delta\varepsilon_{i-1}$ and $\Delta\varepsilon_{i}$,
of the two GB segments connected to node $i$: 
\begin{equation}
u_{i}^{\pm}=\mp Fa+\Delta\varepsilon_{i-1}+\Delta\varepsilon_{i}.\label{eq:u_ij}
\end{equation}
As discussed in Part I \citep{Mishin_2023_RW_part_I}, the exponential
terms in Eq.(\ref{eq:1.02-1}) ensure that the barriers decrease with
increasing force $-u_{i}^{\pm}/a$ but never become strictly zero.
Previous models \citep{Cahn:2001wh,Cottrell:2002ub,Ivanov08a,Chachamovitz:2018vm}
assumed that a barrier could be suppressed to zero at a critical value
of the force. In the present model, the zero-barrier point is regularized
by replacing it with an exponential decay as the force increases. 

The variable $E_{ti}$ in Eq.(\ref{eq:1.02-1}) is the unbiased (when
$u_{i}^{\pm}=0$) jump barrier. In the absence of pinning, the unbiased
barrier is $E_{0}$, which is a model parameter. The unpinned and
unbiased residence time of any GB node is
\begin{equation}
t_{0}=\dfrac{1}{2\nu_{0}}\exp\left(\dfrac{E_{0}}{k_{B}T}\right).\label{eq:1.06a}
\end{equation}
The pinning raises the unbiased jump barrier to $E_{ti}>E_{0}$. Accordingly,
the biased jump barriers $E_{i}^{\pm}$ given by Eq.(\ref{eq:1.02-1})
also increase. In the KMC simulations, the increase of the barriers
due to the pinning effect is implemented by the following algorithm.
After arriving at the current state $i$, the GB node attempts to
make a new jump. After each unsuccessful attempt, we penalize the
node by increasing the jump barriers for both escape routes from the
state $i$. After $n$ unsuccessful attempts, the unbiased barrier
becomes
\begin{equation}
E_{ti}=E_{0}\left(1+(\alpha-1)\dfrac{\sqrt{t/t_{p}}}{1+\sqrt{t/t_{p}}}\right),\label{eq:activation_with_pinning-1}
\end{equation}
where $t=n/2\nu_{0}$ is the discrete time variable. Here, $\alpha>1$
is the pinning strength coefficient and $t_{p}>0$ is the pinning
time, both model parameters. After the node finally makes a successful
jump, the attempt counter $n$ is reset to zero and the process repeats
from the new state. When $t\ll t_{p}$, the barriers grow with time
as $\sqrt{t}$. The square root time dependence reflects the diffusion
kinetics of the solute supply to the GB, causing its pinning. If a
successful jump takes a long time $t\gg t_{p}$, the barrier plateaus
at $E_{ti}=\alpha E_{0}\equiv E_{\infty}>E_{0}$. This long-time limit
represents the saturation of the segregation atmosphere. Once the
atmosphere is saturated, the GB displacements are controlled by the
fully pinned barrier $E_{\infty}$. The most interesting and complex
is the intermediate kinetic regime in which $t_{p}$ is close to the
escape time $t_{0}$. We refer to this kinetic regime as \emph{active
pinning}.

The process described above was implemented in KMC simulations. At
each KMC step, three random numbers $(r_{1},r_{2},r_{3})$ are drawn
uniformly from a unit interval. The first number $r_{1}$ chooses
a GB node, say $i$, for an attempt. All nodes can be chosen with
equal probability. Then $r_{2}$ chooses between a forward jump or
a backward jump, also with equal probability. Finally, $r_{3}$ decides
if the attempt is successful according to the jump probability $P_{i}^{\pm}$.
If the attempt fails, the counter of unsuccessful attempts at site
$i$ is advanced by $1$ and $E_{ti}$ is raised according to Eq.(\ref{eq:activation_with_pinning-1}).
If the jump is successful, the node $i$ is shifted by $\pm a$ along
the $z-$axis, the counter of failed attempts is reset to 0, and the
jump probabilities at nodes $i-2,$ $i-1$, $i$, $i+1$, and $i+2$
are updated. After $2N$ KMC attempts, the clock is advanced by $\nu_{0}^{-1}$.
Other details of the KMC algorithm were discussed in Part I \citep{Mishin_2023_RW_part_I}.

\section{Equilibrium grain boundary properties\label{sec:Equilibrium}}

We will first investigate GB properties in the absence of external
forces ($F=0$). The GB is then only subject to equilibrium thermal
fluctuations. Analysis of this case will create a baseline for comparison
with moving GBs discussed later in section \ref{subsec:Interface-dynamics}.
In addition, since the present SOS model is distinct from previous
versions, a detailed characterization of equilibrium GB properties
will inform future applications of the model

\subsection{Theoretical background}

Without external forces, the average GB position
\begin{equation}
z_{*}=\dfrac{1}{N}\sum_{i=0}^{N-1}z_{i}\label{eq:2.05}
\end{equation}
executes an unbiased random walk while the GB shape fluctuates due
to energy exchanges with the thermostat. The GB properties can be
characterized by the following quantities:
\begin{itemize}
\item Excess GB energy 
\begin{equation}
\overline{\varepsilon}\equiv\dfrac{1}{N}\overline{\left(\sum_{i=0}^{N-1}\varepsilon_{i}\right)}\label{eq:2.03}
\end{equation}
and the mean squared excess GB energy
\begin{equation}
\overline{\varepsilon^{2}}\equiv\dfrac{1}{N}\overline{\left(\sum_{i=0}^{N-1}\varepsilon_{i}^{2}\right)},\label{eq:2.03-1}
\end{equation}
where the bar indicates averaging over a long time. 
\item GB heat capacity per node computed from the energy fluctuation formula
\begin{equation}
C=N\dfrac{\overline{\varepsilon^{2}}-\overline{\varepsilon}^{2}}{k_{B}T^{2}}.\label{eq:1.08-1}
\end{equation}
Note that the GB heat capacity can also be calculated directly by
$C=d\overline{\varepsilon}/dT$.
\item Excess GB area $\bar{s}\equiv\bar{\varepsilon}/\gamma$.
\item Mean squared GB width
\begin{equation}
\overline{w^{2}}\equiv\dfrac{1}{N}\overline{\left(\sum_{i=0}^{N-1}w_{i}^{2}\right)},\label{eq:2.04}
\end{equation}
where
\begin{equation}
w_{i}\equiv z_{i}-z_{*}.\label{eq:2.06}
\end{equation}
\item The GB ``flatness'' parameter $f$ defined as the fraction of parallel
segments ($z_{i}=z_{i+1}$) relative to the total number of segments.
\item Energy self-correlation function
\begin{equation}
K(t)=\dfrac{\left\langle \varepsilon(t)\varepsilon(0)\right\rangle -\overline{\varepsilon}^{2}}{\overline{\varepsilon^{2}}-\overline{\varepsilon}^{2}},\label{eq:1.09}
\end{equation}
where $\varepsilon(t)$ is the instantaneous interface excess energy
per node and the angular brackets indicate averaging over initial
states ($t=0$) along a long simulation trajectory.
\end{itemize}
The GB structure is expected to be nearly planar with a small number
of thermal kinks when $k_{B}T\ll\gamma a^{2}$ and rugged and wavy
when $k_{B}T\gg\gamma a^{2}$. A transition from the first structure,
called ``smooth'', to the second one, called ``rough'', can be expected
to occur when $\gamma a^{2}$ is comparable to $k_{B}T$. 

\begin{figure}
\noindent \begin{centering}
\includegraphics[width=0.8\textwidth]{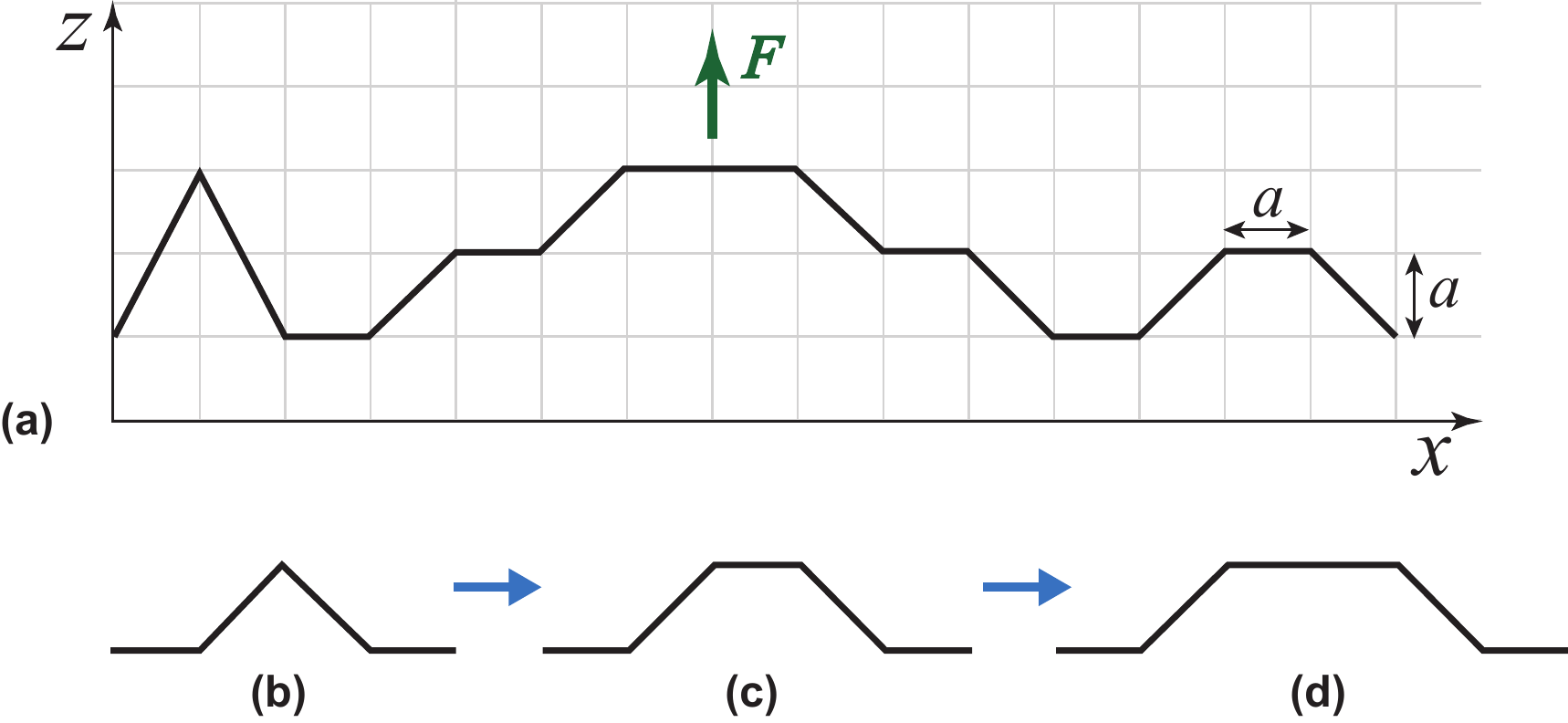}
\par\end{centering}
\caption{(a) 2D GB composed of straight segments connecting nodes of an $a\times a$
square grid. The GB can migrate under an applied force $F$ by stochastic
displacements of the nodes by $\pm a$ at a time parallel to the $z$-axis.
(b) Elementary excitation of a planar GB. (c) The excitation grows
when a neighboring node makes a jump forward, forming a double kink.
(d) Another jump causes further growth of the kink pair.\label{fig:2D-interface}}

\end{figure}

A smooth GB contains kinks as thermal excitations of the perfectly
planar structure. In the present model, the lowest-energy excitation
is the triangular bump shown in Fig.~\ref{fig:2D-interface}(b).
Its excess energy is 
\begin{equation}
u_{2k}=2u_{k}=2(\sqrt{2}-1)\gamma a^{2}\label{eq:1.10}
\end{equation}
and the formation barrier in the absence of pinning is
\begin{equation}
E_{2k}=u_{2k}+E_{0}\exp\left(-\dfrac{u_{k}}{E_{0}}\right),\label{eq:1.11}
\end{equation}
where $u_{k}=(\sqrt{2}-1)\gamma a^{2}$ is the single-kink energy.
One of the base nodes of the triangular bump can jump forward to form
a double kink (Fig.~\ref{fig:2D-interface}(c)). The barrier of this
jump is $E_{0}<E_{2k}$ and the GB energy does not change. The next
jump, shown in Fig.~\ref{fig:2D-interface}(d), causes further separation
of the kinks; it has the same barrier $E_{0}$ and does not change
the GB energy either. Thus, the triangular bump is the critical nucleus
of the kink pair formation. Since the kink pair nucleation barrier
$E_{2k}$ is higher than the kink migration barrier $E_{0}$, the
kink pair formation at a smooth GB is a nucleation-controlled process.

The thermal kink concentration (probability per node) on a smooth
GB is \citep{Hirth,Saito:1996vo}
\begin{equation}
n_{k}=2\exp\left(-\dfrac{u_{k}}{k_{B}T}\right),\label{eq:1.11.1}
\end{equation}
from which the excess GB energy is
\begin{equation}
\overline{\varepsilon}=2u_{k}\exp\left(-\dfrac{u_{k}}{k_{B}T}\right).\label{eq:1.11.2}
\end{equation}
The GB heat capacity calculated from this energy,
\begin{equation}
C=\dfrac{d\overline{\varepsilon}}{dT}=\dfrac{2u_{k}^{2}}{k_{B}T^{2}}\exp\left(-\dfrac{u_{k}}{k_{B}T}\right),\label{eq:1.11.3}
\end{equation}
reaches a maximum at $k_{B}T/\gamma a^{2}=(\sqrt{2}-1)/2\approx0.21$.
This maximum can be associated with the GB roughening transition.
Note that the kink concentration corresponding to this maximum is
$n_{k}=2e^{-2}\approx0.27$, which is no longer small.

Above the roughening transition, the GB develops significant shape
fluctuations and can be better described by the capillary wave theory
\citep{Gelfand:1990vz,Lapujoulade:1994uw,Saito:1996vo}. For 2D interfaces,
the capillary wave amplitude diverges to infinity with increasing
lateral size $L=Na$. The relevant results of the theory are summarized
in Appendix A. The theory predicts the mean squared GB width 
\begin{equation}
\overline{w^{2}}=\dfrac{k_{B}TL}{12\Lambda a},\label{eq:3.12-1}
\end{equation}
where 
\begin{equation}
\Lambda=f_{0}+(\partial^{2}f/\partial\beta^{2})_{0}\label{eq:3.2-1}
\end{equation}
is the GB stiffness, $f$ is the GB free energy per unit area, and
$\beta$ is the small angle between the local GB orientation and the
$x$-axis. The first term in the right-hand side of Eq.(\ref{eq:3.2-1})
is the GB free energy in the $\beta\rightarrow0$ limit (perfectly
planar GB). The second (``torque'') term is the second angular derivative
of $f$ taken in the $\beta\rightarrow0$ limit. It should be emphasized
that the underlying assumption of the theory is that $\beta\ll1$,
i.e., $\sqrt{\overline{w^{2}}}\ll L$.

\subsection{Simulation results}

The KMC simulations were performed in normalized variables obtained
by dividing the time, the coordinates, and all energies by $\nu_{0}^{-1}$,
$a$, and $E_{0}$, respectively. The normalized temperature, force,
and GB energy become, respectively,
\[
\theta=\dfrac{k_{B}T}{E_{0}},\enskip\varphi=\dfrac{Fa}{E_{0}},\enskip\sigma=\dfrac{\gamma a^{2}}{E_{0}}.
\]
Other normalized variables used in the simulations are summarized
in Table \ref{tab:Reduced-variables}. The results presented in this
subsection were obtained at $\sigma=1$.

\subsubsection{GB properties in the absence of pinning}

We first consider the simulation results in the absence of pinning.
Recall that the GB is not acted upon by any external force ($\varphi=0$).

\begin{figure}
\textbf{(a)} \includegraphics[width=0.8\textwidth]{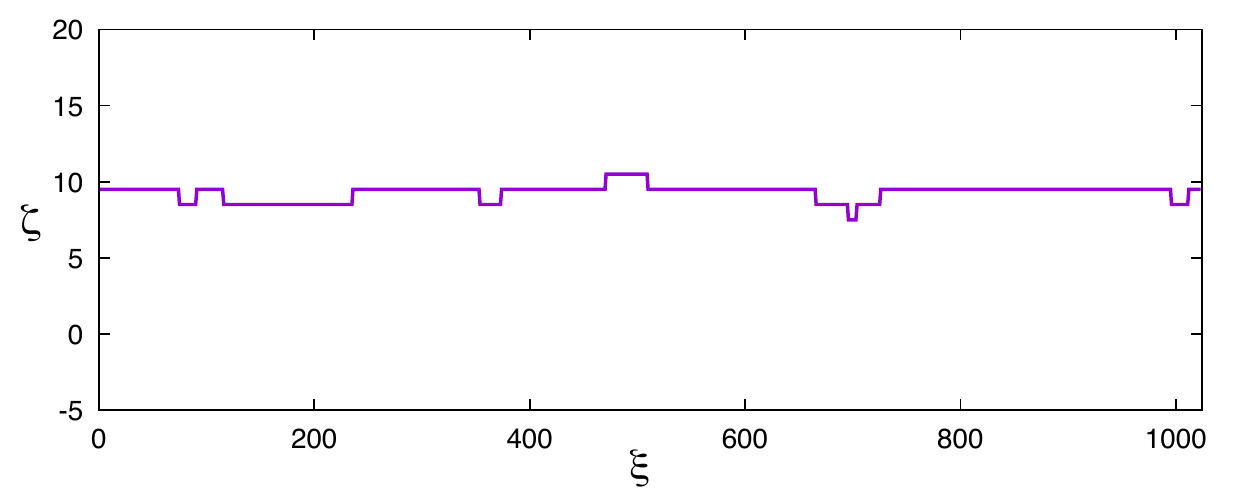}

\bigskip{}
\bigskip{}
\bigskip{}

\textbf{(b)} \includegraphics[width=0.8\textwidth]{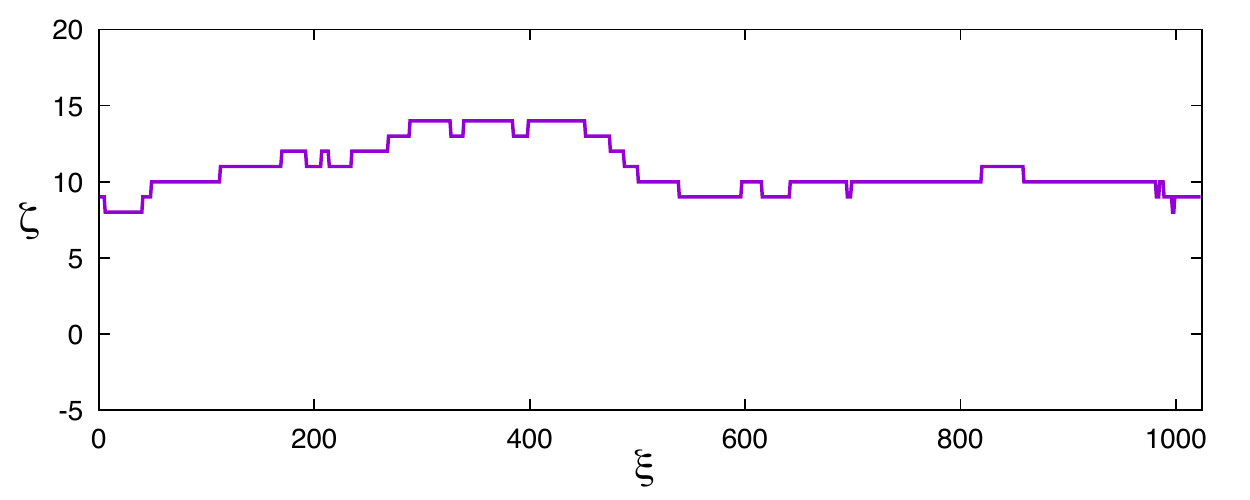}

\bigskip{}
\bigskip{}
\bigskip{}

\textbf{(c)} \includegraphics[width=0.8\textwidth]{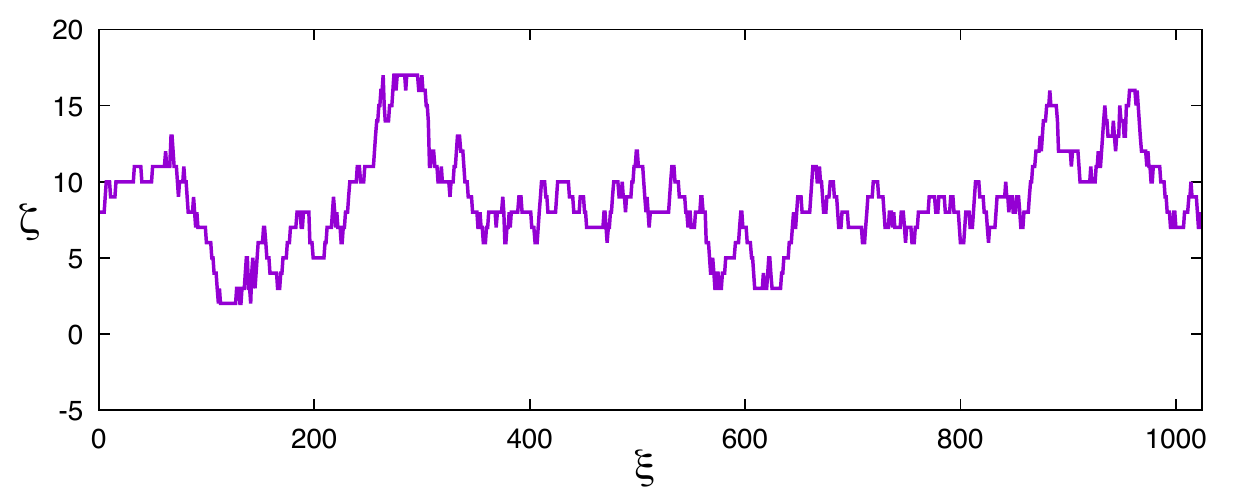}

\caption{Typical interface shapes at $\sigma=1$. (a) Smooth interface ($\theta=0.083$);
(b) Moderately rough interface ($\theta=0.095$); (c) Fully rough
interface ($\theta=0.285$). The GB is composed of $N=1024$ nodes.
\label{fig:Interface-shapes.}}
\end{figure}

Figure \ref{fig:Interface-shapes.} shows typical structures of a
smooth GB with a small concentration of kinks, a rough GB with capillary
waves, and a moderately rough GB in between. As expected, the GB evolves
from smooth to rough with increasing temperature. 

\begin{figure}
\begin{centering}
\includegraphics[width=0.7\textwidth]{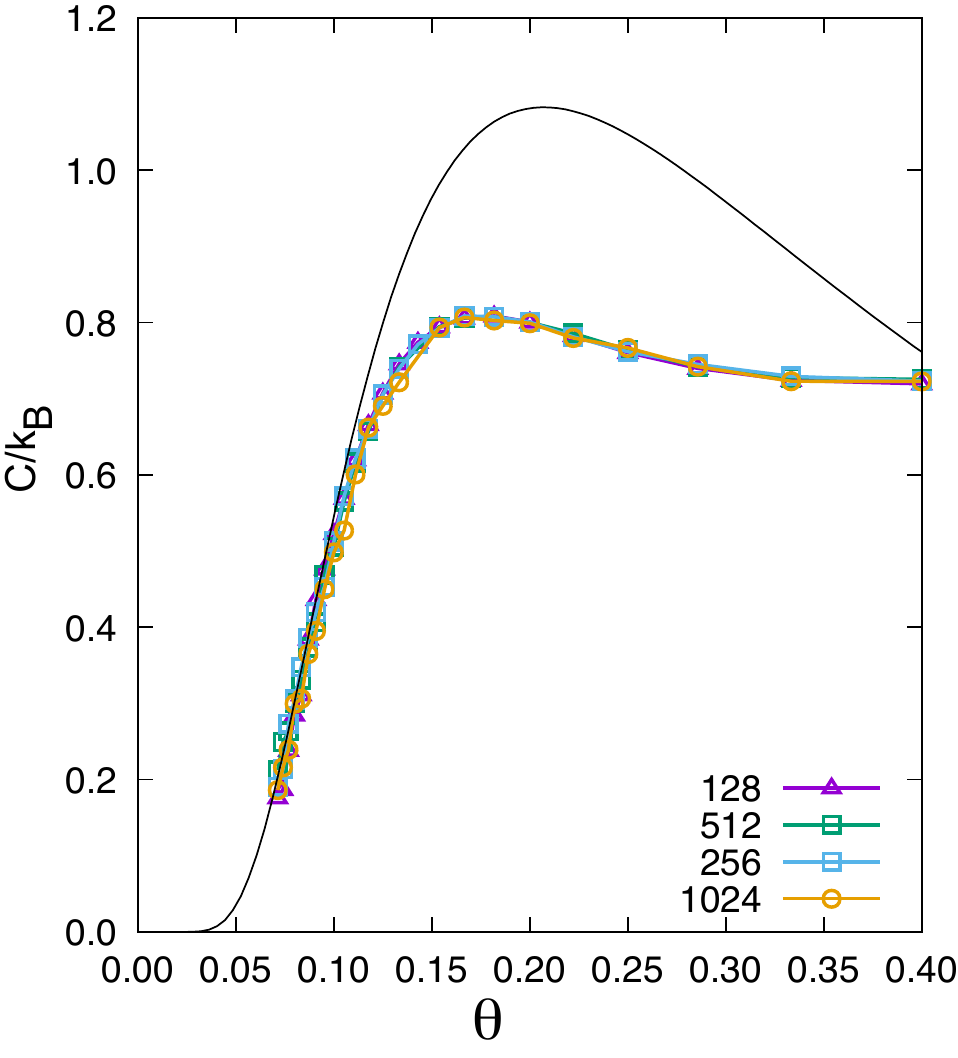}
\par\end{centering}
\caption{GB heat capacity as a function of temperature $\theta$ at $\sigma=1$
for several system sizes $N$ indicated in the key. The boundary is
not subject to external forces or solute pinning. As $N$ increases
by nearly an order of magnitude, the heat capacity curves coincide
within the statistical scatter. The solid curve is predicted by Eq.(\ref{eq:1.11.3})
based on the kink model. \label{fig:Interface-heat-capacity}}
\end{figure}

To understand the nature of the roughening transition, we examine
the temperature dependence of the GB heat capacity computed from the
fluctuation formula (\ref{eq:1.08-1}). Figure \ref{fig:Interface-heat-capacity}
shows that the heat capacity obtained by the simulations reaches a
maximum when $\theta/\sigma$ is reasonably close to 0.2, as predicted
by the kink model mentioned above. The simulations accurately follow
Eq.(\ref{eq:1.11.3}) at low temperatures when the GB is fairly smooth.
The agreement worsens with temperature, and the maximum predicted
by Eq.(\ref{eq:1.11.3}) significantly overshoots the simulation results.
This is unsurprising given that the GB structure near the maximum
is intermediate between smooth and rough, and the kink model is a
crude approximation. For a true phase transformation, the height of
the heat capacity maximum must increase with the system size and diverge
to infinity in the thermodynamic limit ($N\rightarrow\infty$). By
contrast, the heat capacity obtained by the simulations is virtually
independent of the system size. Thus, in the present model, the GB
roughening is a \emph{continuous} transformation. Continuous roughening
was also predicted by other models of 2D interfaces \citep{Lapujoulade:1994uw,LandauBinder09,Saito:1996vo}. 

\begin{figure}
\textbf{(a)} \includegraphics[width=0.5 \textwidth]{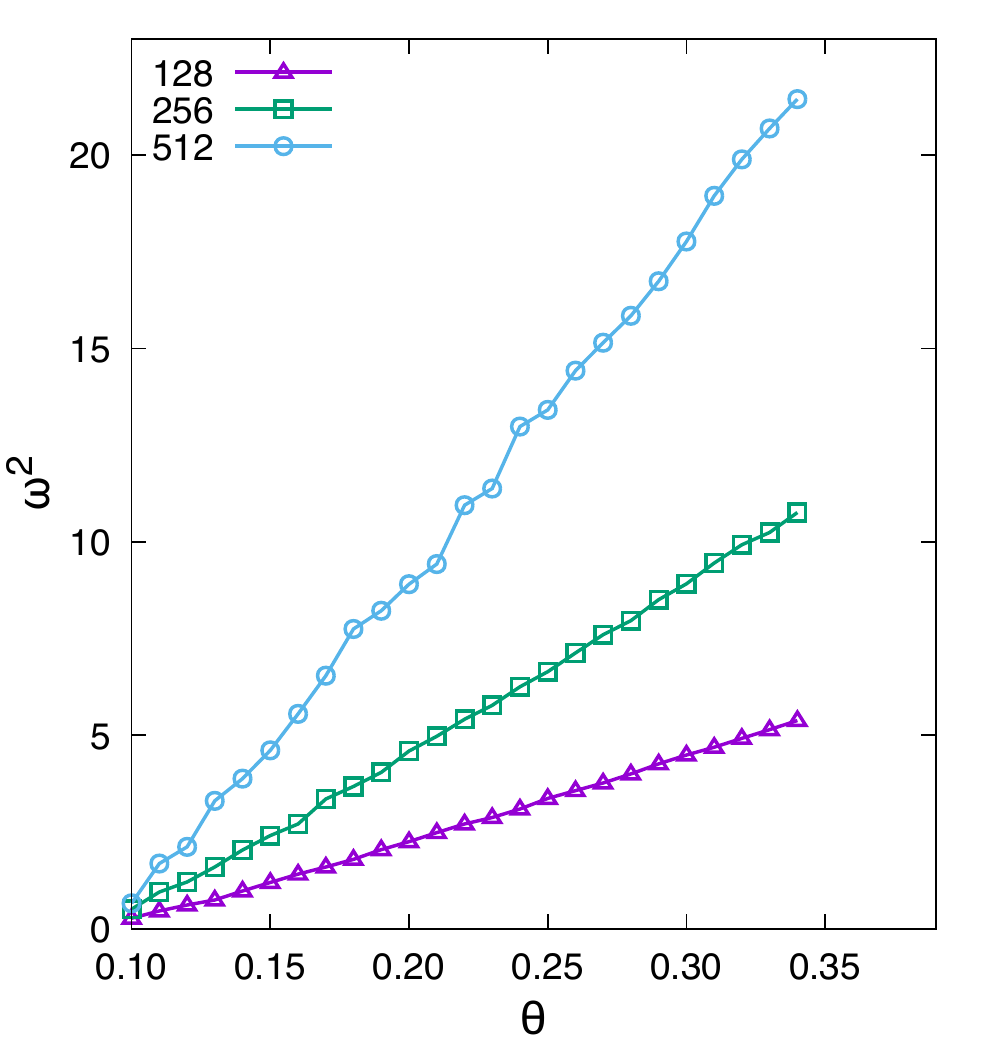}

\bigskip{}
\bigskip{}

\textbf{(b)} \includegraphics[width=0.52 \textwidth]{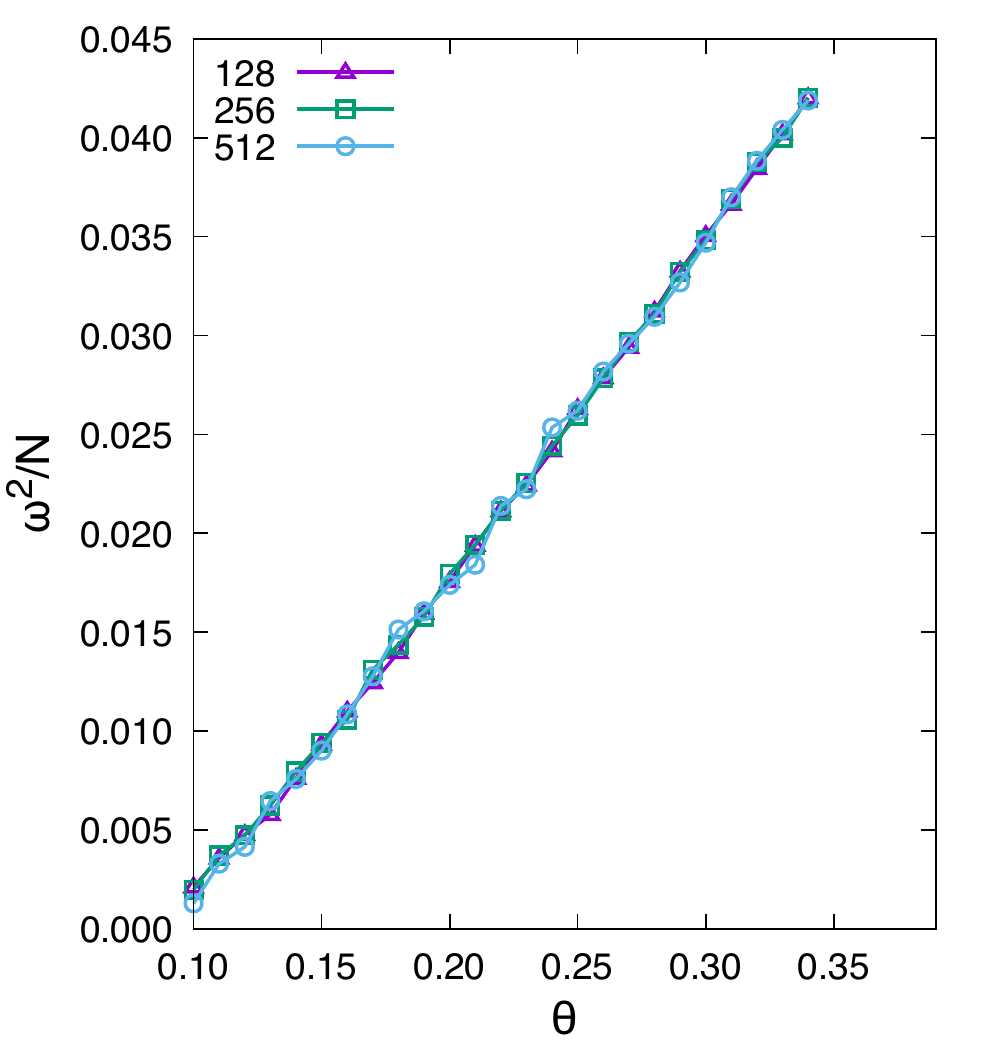}

\caption{(a) Mean squared GB width $\overline{\omega^{2}}$ as a function of
temperature $\theta$ at $\sigma=1$ for three system sizes $N$ indicated
in the key. The monotonic increase of $\overline{\omega^{2}}$ with
$N$ proves the GB roughness. (b) The plots of $\overline{\omega^{2}}/N$
versus $\theta$ for different $N$ values collapse into a single
curve, confirming the size scaling predicted by Eq.(\ref{eq:3.12-1}).\label{fig:Interface-width}}
\end{figure}

\begin{figure}
\noindent \begin{centering}
\includegraphics[width=0.7 \textwidth]{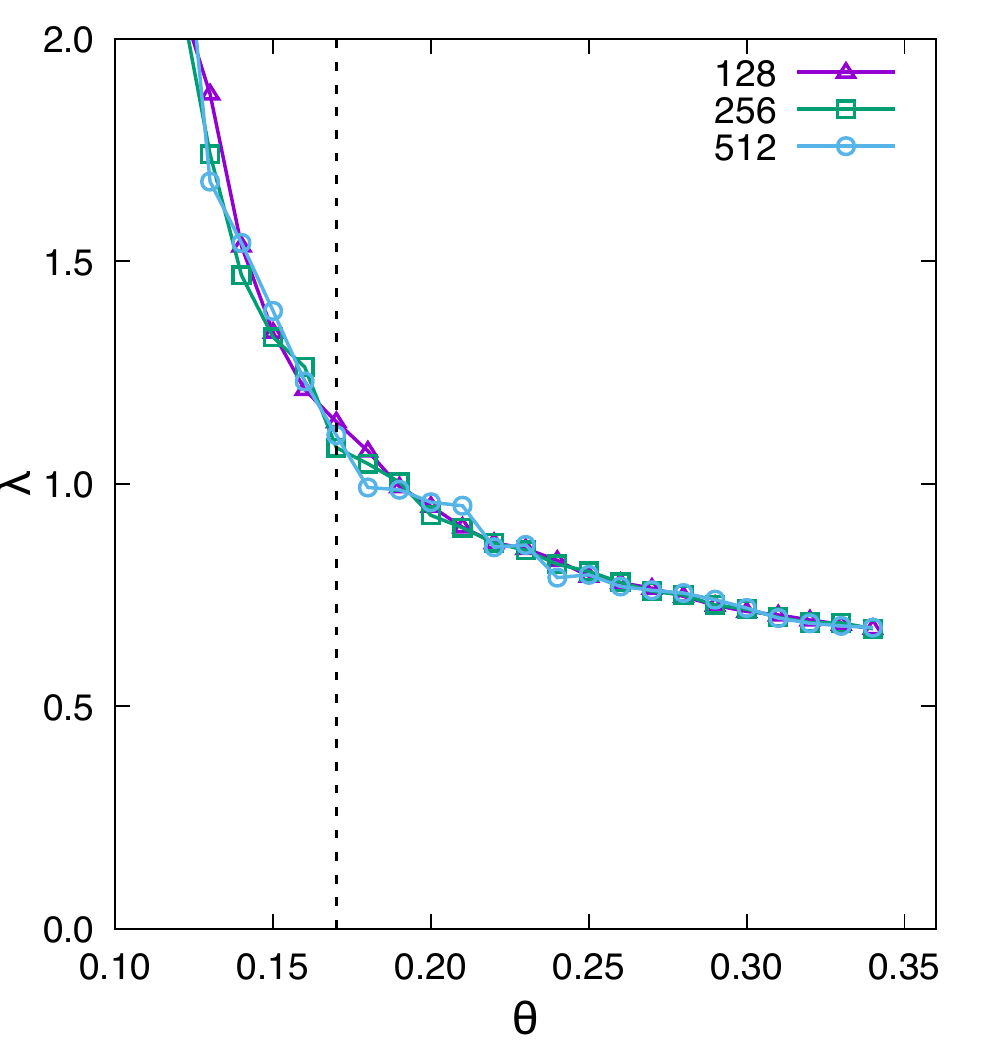}
\par\end{centering}
\caption{GB stiffness $\lambda$ as a function of temperature computed from
Eq.(\ref{eq:2-3-1}) at $\sigma=1$ for three different system sizes
$N$. The dashed line marks the approximate location of the GB roughening
transition. \label{fig:Stiffness}}

\end{figure}

In the capillary wave regime, Eq.(\ref{eq:3.12-1}) predicts that
the mean squared GB width $\overline{\omega^{2}}$ increases with
the system size $N$ and diverges to infinity at $N\rightarrow\infty$.
This divergence is considered a formal definition of a rough interface.
The expected increase of $\overline{\omega^{2}}$ with $N$ is indeed
observed in the simulations (Fig.~\ref{fig:Interface-width}(a)),
confirming that the GB is officially rough above $\theta\approx0.2$.
Furthermore, the plots of $\overline{\omega^{2}}/N$ versus $\theta$
for different $N$ values collapse into a single master curve (Fig.~\ref{fig:Interface-width}(b))
as predicted by Eq.(\ref{eq:3.12-1}). This curve allows us to extract
the normalized interface stiffness $\lambda$ using the dimensionless
form of Eq.(\ref{eq:3.12-1}):
\begin{equation}
\lambda=\dfrac{\theta}{12\left(\overline{\omega^{2}}/N\right)}.\label{eq:2-3-1}
\end{equation}
The stiffness obtained from this equation is plotted as a function
of temperature in Fig.~\ref{fig:Stiffness}. The sharp increase in
$\lambda$ below the roughening transition is an artifact because
the capillary wave model is only valid for rough interfaces. The fact
that $\lambda$ decreases with increasing temperature points to a
significant contribution of the configurational entropy to the GB
free energy associated with the GB shape fluctuations. 

\begin{figure}
\begin{centering}
\textbf{(a)} \includegraphics[width=0.5 \textwidth]{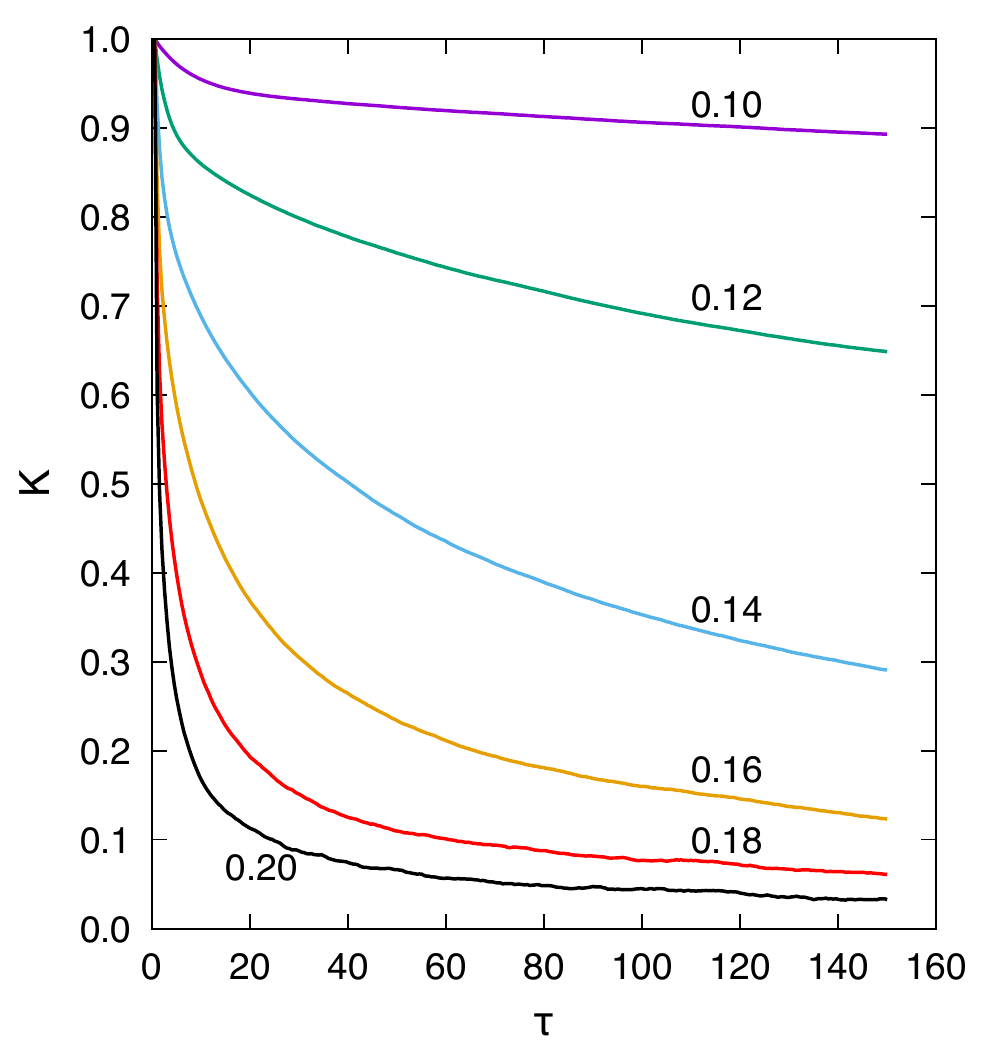}
\par\end{centering}
\bigskip{}
\bigskip{}
\bigskip{}

\begin{centering}
\textbf{(b)} \includegraphics[width=0.5 \textwidth]{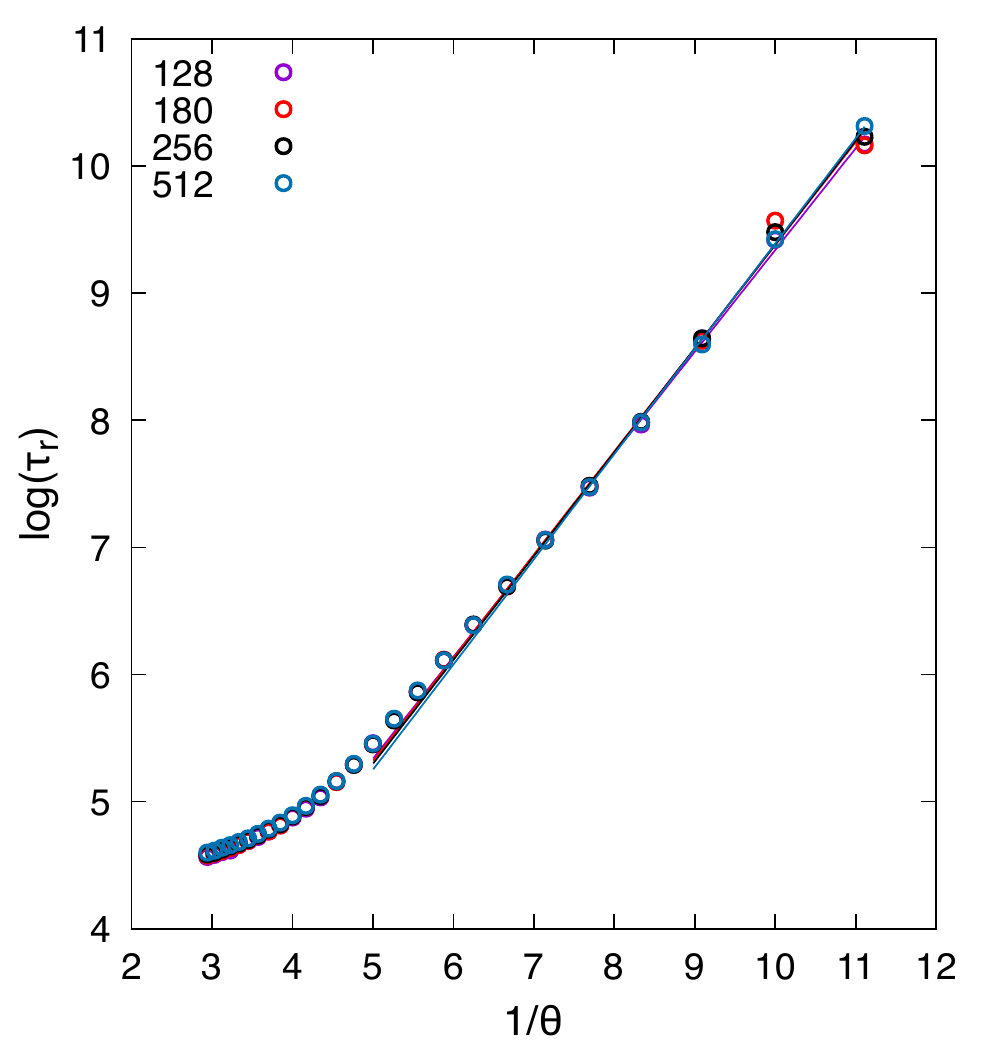}
\par\end{centering}
\caption{(a) Energy self-correlation function $K(\tau)$ for $N=256$ and $\sigma=1$
at several temperatures $\theta$ indicated in the labels. (b) Arrhenius
diagram of the energy relaxation time $\tau_{r}$ for different system
sizes indicated in the key. The straight lines are linear fits to
the low-temperature portions of the curves.  \label{fig:(a)-Energy-correlation}}
\end{figure}

Figure \ref{fig:(a)-Energy-correlation}(a) illustrates typical energy
self-correlation functions $K(\tau)$ at several temperatures including
both smooth and rough GB structures. The correlation decay rate increases
with temperature, as it should for a thermally activated process.
The decay rate can be quantified by extracting the relaxation time
$\tau_{r}$ by fitting the short-time portion of $K(\tau)$ with the
exponential relation $K=K_{0}\exp(-\tau/\tau_{r})$, where $K_{0}$
is a constant. The relaxation times obtained are plotted in the Arrhenius
coordinates $\ln\tau_{r}$ versus $1/\theta$ in Fig.~\ref{fig:(a)-Energy-correlation}(b).
Observe that the curves corresponding to different system sizes coincide
within the scatter of the points, confirming the local nature of the
short-time relaxation. The Arrhenius plots are fairly linear at low
temperatures but develop a significant upward deviation above the
roughening transition. The low-temperature portions of the curves,
corresponding to smooth GB structures, were fitted with the Arrhenius
relation
\begin{equation}
\tau_{r}=\tau_{r}^{0}\exp\left(-\dfrac{\varepsilon_{a}}{\theta}\right),\label{eq:2-3-2}
\end{equation}
where $\tau_{r}^{0}$ is a constant. The activation energies $\varepsilon_{a}$
extracted from the fits were found to be practically the same for
all system sizes and equal to $\varepsilon_{a}=0.821$. This number
is reasonably close to the normalized kink pair energy $u_{2k}/E_{0}=2(\sqrt{2}-1)\sigma=0.828$,
see Eq.(\ref{eq:1.10}) above and recall that $\sigma=1$ in the simulations.
This agreement confirms that the dominant relaxation mechanism in
a smooth GB is the nucleation and recombination of thermal kink pairs.
The non-Arrhenius deviation at high temperatures reflects the gradual
transition from the smooth to the rough GB structure when the fluctuations
form capillary waves. Under such conditions, the relaxation mechanism
involves collective rearrangements of the high concentration of geometrically
necessary kinks accommodating the GB curvature. Such rearrangements
occur on a greater length scale than in a smooth GB, causing the significant
increase in the relaxation time. 

\subsubsection{GB properties with pinning\label{GB-properties}}

The effect of pinning on the equilibrium GB properties is controlled
by the relative pinning time $\tau_{p}/\tau_{0}$, where $\tau_{0}=t_{0}\nu_{0}$
is the unpinned residence time. To investigate the pinning effect,
the KMC simulations were performed at several fixed values of $\tau_{p}/\tau_{0}$
spanning the range from 0.01 to 100. As temperature was increased,
$\tau_{0}$ decreased (cf.~Eq.(\ref{eq:1.06a})) but $\tau_{p}$
was adjusted to keep $\tau_{p}/\tau_{0}$ constant. The GB energy
and the pinning strength were fixed at $\sigma=1$ and $\alpha=1.5$,
respectively. 

\begin{figure}
\noindent \begin{centering}
(a) \includegraphics[width=0.55 \textwidth]{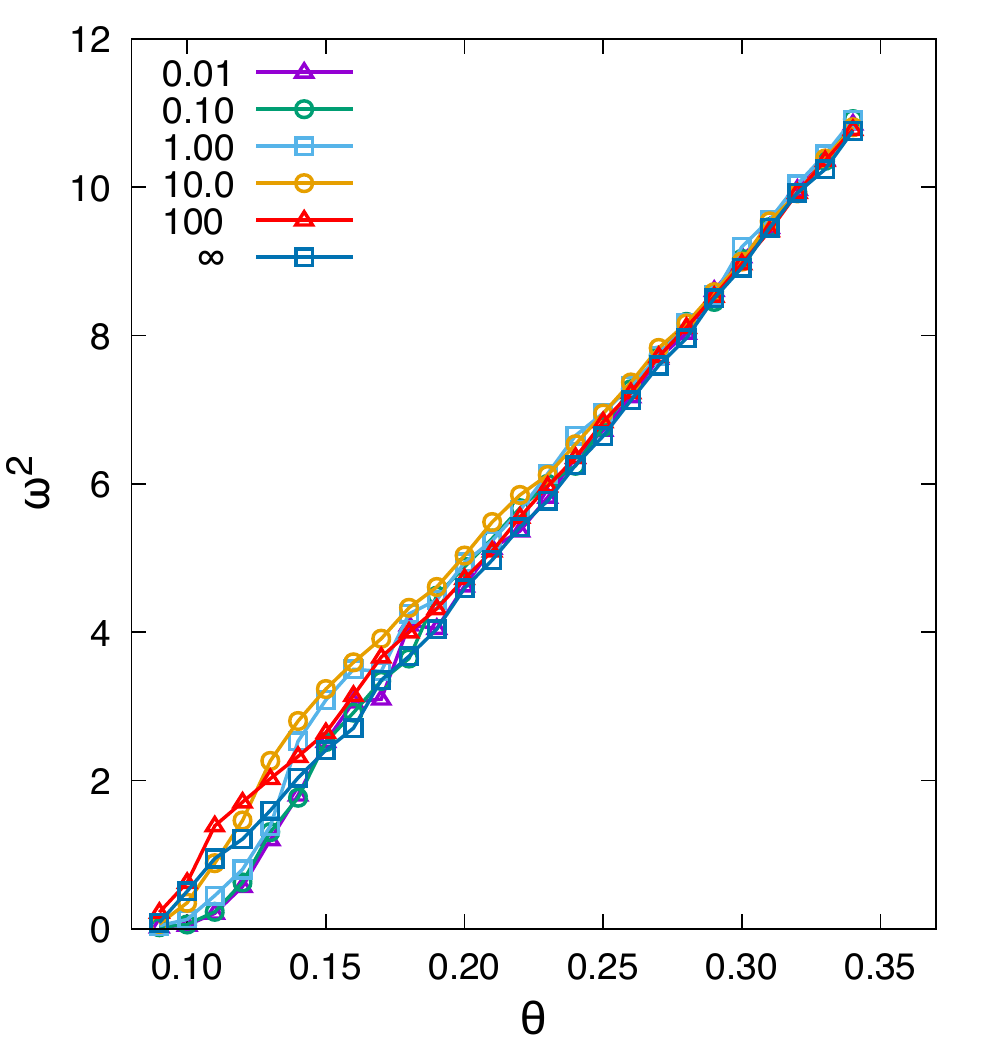}
\par\end{centering}
\bigskip{}
\bigskip{}

\noindent \begin{centering}
(b) \includegraphics[width=0.55 \textwidth]{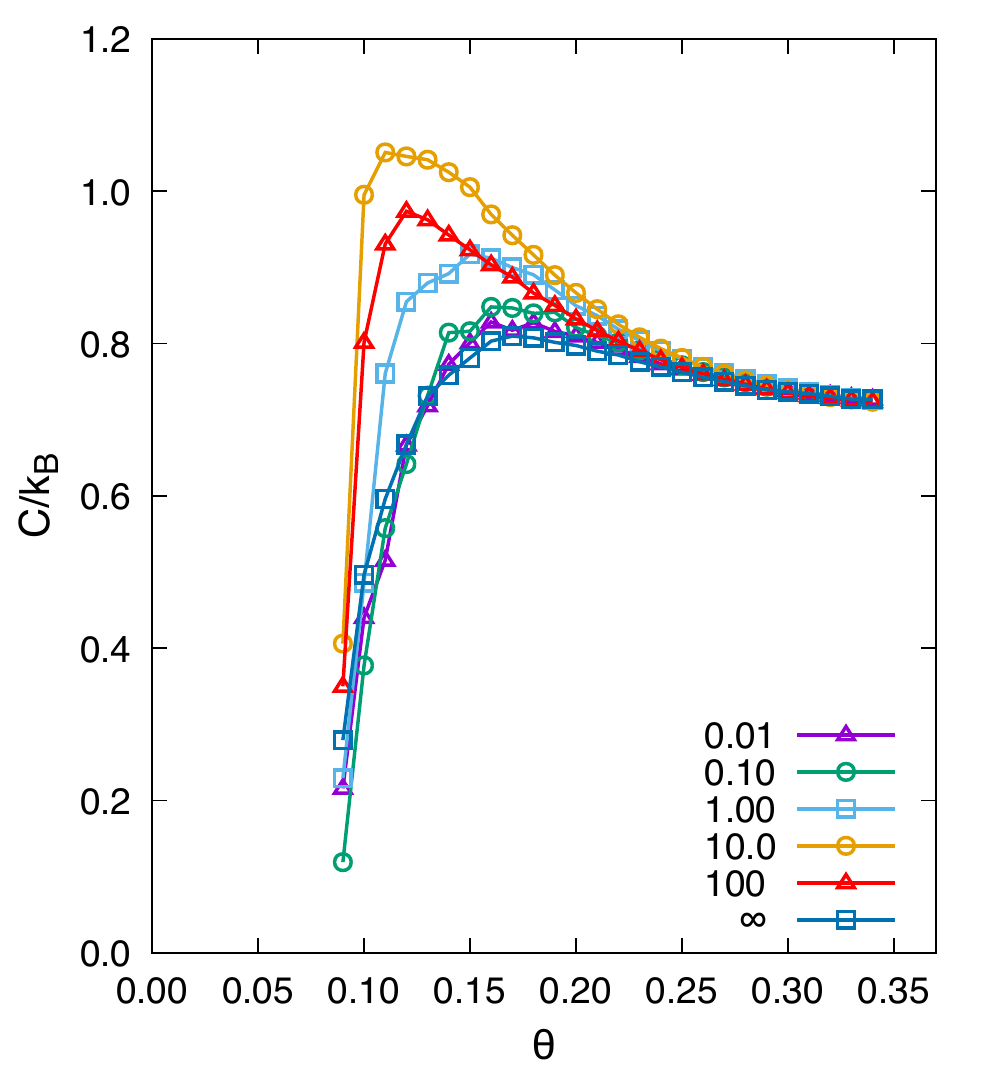}
\par\end{centering}
\caption{Effect of pinning on GB properties without driving forces. The KMC
simulations were performed at $N=256$, $\sigma=1$, and several values
of the relative pinning time \emph{$\tau_{p}/\tau_{0}$} indicated
in the key. The curve labeled $\infty$ corresponds to the absence
of pinning. (a) Mean squared GB width $\overline{\omega^{2}}$ as
a function of temperature $\theta$. (b) GB heat capacity as a function
of temperature $\theta$.\label{fig:Effect-of-pinning} }

\end{figure}

The general trend observed in the simulations is that the active pinning
promotes GB roughness. For example, Figure \ref{fig:Effect-of-pinning}(a)
presents the mean squared GB width $\overline{\omega^{2}}$ plotted
as a function of temperature for a set of $\tau_{p}/\tau_{0}$ values.
At temperatures well above the roughening transition, $\overline{\omega^{2}}$
is practically independent of $\tau_{p}/\tau_{0}$ and increases as
a linear function of temperature, as predicted by the capillary-wave
equation (\ref{eq:3.12-1}). Thus, the pinning has little effect on
the rough GB structure. As temperature decreases and the GB becomes
smoother, $\overline{\omega^{2}}$ deviates from the linear behavior.
At large and small $\tau_{p}/\tau_{0}$ values, $\overline{\omega^{2}}$
approaches the simulation results obtained in the absence of pinning
(the curve labeled $\infty$). At intermediate $\tau_{p}/\tau_{0}$
values corresponding to the active pinning regime (e.g., $\tau_{p}/\tau_{0}$
between 1 and 10), $\overline{\omega^{2}}$ displays upward deviations
that grow as temperature decreases. The GB becomes wider and thus
rougher compared with the unpinned and fully pinned cases. At even
low temperatures, $\overline{\omega^{2}}$ converges to zero as the
GB attains a smooth structure. In other words, the pinning effect
on the GB width is strongest at temperatures near the roughening transition
and when the pinning time $\tau_{p}$ is on the order of $\tau_{0}$.

These observations are consistent with the temperature dependence
of the heat capacity shown Figure \ref{fig:Effect-of-pinning}(b).
It should be reminded that the fluctuation formula (\ref{eq:1.08-1})
does not necessarily give the correct heat capacity under the active
pinning conditions because the KMC simulations become non-Markovian
\citep{Mishin_2023_RW_part_I}. Nevertheless, the heat capacity obtained
from Eq.(\ref{eq:1.08-1}) can be used as simply a measure of energy
fluctuations, which are expected to grow near the roughening transition.
As with the mean squared GB width, at temperatures above the roughening
transition, the heat capacity is unaffected by pinning: the results
obtained at all $\tau_{p}/\tau_{0}$ values converge to the same curve
obtained by unpinned simulations. At large and small $\tau_{p}/\tau_{0}$
values, the heat capacity continues to follow the unpinned curve.
However, in the active pinning regime, the heat capacity displays
significant upward deviations at temperatures close to the roughening
transition. Taking the peak position as the transition temperature,
we observe that active pinning shifts the roughening transition temperature
down and makes the transition sharper. 

\section{Grain boundary dynamics\label{subsec:Interface-dynamics}}

\subsection{GB dynamics in the absence of pinning\label{subsec:GB-dynamics}}

Now suppose that the GB is acted upon by an external force $F$. We
first disregard the pinning effect and consider the motion of a chemically
pure GB. 

\begin{figure}
\begin{centering}
\includegraphics[width=0.7\textwidth]{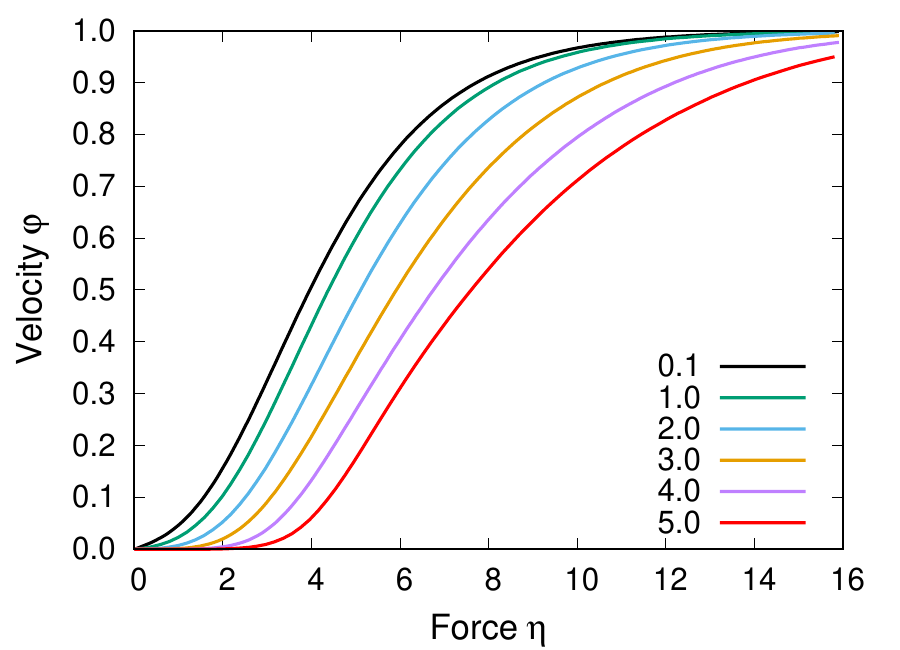}
\par\end{centering}
\caption{Velocity-force relations in the 2D GB model with $N=512$ nodes at
the temperature of $\theta=0.2$ for several values of the GB energy
$\sigma$ indicated in the key. The GB is not subject to pinning.\label{fig:Velocity-force-relations}}

\end{figure}

Examples of velocity-force functions obtained by the simulations are
shown in Fig.~\ref{fig:Velocity-force-relations} for several values
of the normalized GB energy $\sigma$. The simulation temperature
is fixed at $\theta=0.2$, which is close to the roughening transition
in a stationary GB. Since large GB energy enforces planar GB shape,
one would expect that, as $\sigma$ increases, the velocity-force
curves should approach those predicted by the 1D model. Contrary to
this expectation, it is the lowest-energy curve ($\sigma=0.1$) that
is practically indistinguishable from the curve computed previously
\citep{Mishin_2023_RW_part_I} within the 1D model (not shown in the
figure). As $\sigma$ increases, the 2D results increasingly deviate
from the 1D model. The curves develop a nearly flat portion at low
velocities, followed by a rapid rise as the driving force increases.
Similar shapes of velocity-force functions were seen in some of the
previous 2D and 3D simulations of interface dynamics \citep{Mendelev02a,Mendelev:2001aa,Mendelev:2001wk,Mendelev:2002wv,Wicaksono:2013aa}.
Note that a stationary GB with $\sigma=0.1$ is rough, a stationary
GB with $\sigma=1$ is transitional between rough and smooth, and
stationary GBs with $\sigma=2$, $3$, $4$ and $5$ are smooth. Two
conclusions can be drawn from these observations:
\begin{itemize}
\item 1D models, including the classical models \citep{Cahn-1962,Lucke-Stuwe-1963,Lucke:1971aa},
should be interpreted as representing driven motion of the average
plane of a \emph{rough} GB.
\item Smooth GBs strongly deviate from the classical models by displaying
much lower mobility. 
\end{itemize}
The mechanism of the latter deviation is as follows. Smooth GBs undergo
a dynamic roughening transition as the velocity increases at a fixed
temperature. This transition alters the migration mechanism and increases
the GB mobility. The migration mechanism is mediated by the motion
of kink pairs when the GB is smooth and by a biased random walk of
the average GB plane when the GB becomes rough. As mentioned above,
a rough GB contains a high concentration of geometrically necessary
kinks accommodating the capillary waves. Their motion is responsible
for both the capillary fluctuations and, in the presence of a driving
force, for the drift of the average GB plane. Due to the large kink
concentration, the GB mobility is high. By contrast, a smooth GB contains
a small concentration of thermal kinks that can only support slow
GB migration. A large enough force can cause the nucleation of additional
(non-equilibrium) kinks and eventually cause a dynamic roughening
transition. The latter, in turn, accelerates the GB migration and
is responsible for the sigma shape of the velocity-force curves at
large $\sigma$ values (Fig.~\ref{fig:Velocity-force-relations}).

The dynamic roughening transition is illustrated Figure \ref{fig:Excess-interface-area}
using the excess area $s$ as a measure of GB roughness. In the stationary
state ($F=0$), the three GBs shown in this plot are, respectively,
smooth ($\sigma=5$), intermediate ($\sigma=1$), and rough ($\sigma=0.1$).
When a force is applied and causes GB motion, the GB roughness increases
with the force. The initially smooth GB develops non-equilibrium kinks
and eventually reaches the excess area characteristic of rough GBs.
The initially rough interface becomes even rougher. Fig.~\ref{fig:Dynamic-roughness}
visually represents the dynamic roughening transition with increasing
velocity. As in the stationary case, this transition is continuous.

We next discuss the kink-mediated migration of a smooth GB in more
detail. Several models of interface migration by the kink pair mechanism
were proposed \citep{Hirth,Bertocci-1969,Seeger-1981,Mendelev:2001wk}.
(We note in passing that the problem is similar to kink-mediated dislocation
glide if elastic effects are neglected.) We assume that the kinks
only nucleate by pairs and their thermally equilibrium concentration
is small. The driving force reduces the barrier of forward GB jumps
and creates a relatively high concentration $n_{2k}$ of non-equilibrium
kink pairs (number per unit area) bounding GB segments displaced in
the force direction. The force also biases the barriers of kink jumps
parallel to the GB plane, driving kink separation in each pair. The
kink pair growth velocity is $2v$, where $v$ is the single-kink
drift velocity along the planar boundary given by
\begin{equation}
v=a\nu_{0}\left[\exp\left(-\dfrac{E^{(+)}}{k_{B}T}\right)-\exp\left(-\dfrac{E^{(-)}}{k_{B}T}\right)\right],\label{eq:1.04}
\end{equation}
where
\begin{equation}
E^{(+)}=E_{0}\exp\left(-\dfrac{Fa}{2E_{0}}\right)\label{eq:1.03a}
\end{equation}
is the forward jumps barrier and 
\begin{equation}
E^{(-)}=Fa+E_{0}\exp\left(-\dfrac{Fa}{2E_{0}}\right)\label{eq:1.03b}
\end{equation}
is the backward jump barrier. The velocity-force relation predicted
by Eq.(\ref{eq:1.04}) is linear under a small force and becomes nonlinear
as the force increases.

Suppose the GB is initially planar and the applied force causes the
nucleation of $J$ kink pairs per unit area per unit time. After a
time $t$, the pair concentration becomes $n_{2k}=Jt$. During this
time, the nuclei grow to the average size $2vt$. The time required
to form a contiguous new layer is found from the condition $2vt=l_{2k}$,
where $l_{2k}=1/n_{2k}a$ is the average distance between the nucleation
centers. This crude estimate gives the time $t=1/\sqrt{2avJ}$ in
which the GB displaces a distance $a$ (one layer thickness) in the
force direction. For the GB velocity $v_{2k}=a/t$ we then have
\begin{equation}
v_{2k}=a\sqrt{2avJ,}\label{eq:GB_vel}
\end{equation}
where the subscript $2k$ indicates that this velocity is specific
to the kink pair mechanism. Equation (\ref{eq:GB_vel}) was previously
derived by Bertocci \citep{Bertocci-1969} by a different method. 

The next step is to calculate the nucleation rate $J$. This rate
is controlled by the nucleation rate of triangular bumps shown in
Fig.~\ref{fig:2D-interface}(b):
\begin{equation}
J=\dfrac{\nu_{0}}{a^{2}}P_{01}P_{s},\label{eq:nucleation}
\end{equation}
where 
\begin{equation}
P_{01}=\exp\left(-\dfrac{u_{2k}-Fa+E_{0}e^{-(u_{2k}-Fa)/2E_{0}}}{k_{B}T}\right)\label{eq:P_01}
\end{equation}
is the probability per one attempt that a given node on a planar GB
(call it state 0) pops up to form a triangular bump (call it state
1). The factor $P_{s}$ in Eq.(\ref{eq:nucleation}) is the probability
of survival of the bump. Indeed, the bump can disappear by a reverse
jump $1\rightarrow0$ whose probability per attempt is
\begin{equation}
P_{10}=\exp\left(-\dfrac{E_{0}e^{-(u_{2k}-Fa)/2E_{0}}}{k_{B}T}\right).\label{eq:P_10}
\end{equation}
Alternatively, the triangular bump can expand into a kink pair containing
two GB nodes (state 2) as shown in Fig.~\ref{fig:2D-interface}(c).
This kink pair can collapse back into the triangular bump (jump $2\rightarrow1$)
or expand further. The kink separation will then execute a driven
random walk in which it will most likely keep expanding indefinitely.
But there is a chance that the kink pair eventually collapses back
into a triangular bump. $P_{s}$ is the probability that such a collapse
does not happen. As shown in Appendix B, 
\begin{equation}
P_{s}=\dfrac{2(P_{12}-P_{21})}{P_{10}+2(P_{12}-P_{21})}.\label{eq:Escape_prob}
\end{equation}
Here,
\begin{equation}
P_{12}=\exp\left(-\dfrac{E_{0}e^{-Fa/2E_{0}}}{k_{B}T}\right)\label{eq:P_12}
\end{equation}
and 
\begin{equation}
P_{21}=\exp\left(-\dfrac{Fa+E_{0}e^{-Fa/2E_{0}}}{k_{B}T}\right)\label{eq:P_21}
\end{equation}
are the $1\rightarrow2$ and $2\rightarrow1$ jump probabilities (per
attempt), respectively. Note that they are related to the single-kink
drift velocity (\ref{eq:1.04}) by
\begin{equation}
v=a\nu_{0}(P_{12}-P_{21}).\label{eq:vel}
\end{equation}
Putting all pieces together, the GB velocity by the kink pair mechanism
becomes
\begin{equation}
v_{2k}=2v\left(\dfrac{P_{01}}{P_{10}+2(P_{12}-P_{21})}\right)^{1/2}.\label{eq:GB_vel-1}
\end{equation}

Figure \ref{fig:kink-mechanism} shows that Eq.(\ref{eq:GB_vel-1})
compares with the KMC simulations reasonably well when the GB energy
is not too high. However, it increasingly overestimates the velocity
as the GB energy increases. This is understandable because the nucleation
rate decreases with increasing GB energy, and the nuclei spacing $l_{2k}$
eventually reaches the system size $L=Na$ ($N=512$ in the simulations).
The GB migration becomes strongly nucleation-controlled. A single
nucleation event triggers rapid expansion of the kink pair to the
system size before another pair has a chance to nucleate. The expectation
time of the nucleation event is $t=(JaL)^{-1}$. Thus, instead of
Eq.(\ref{eq:GB_vel}), the GB velocity becomes 
\begin{equation}
v_{2k}=a/t=Ja^{2}L.\label{eq:GB_vel_2}
\end{equation}
Note that Eq.(\ref{eq:GB_vel}) can be written in the form $v_{2k}=Ja^{2}l_{2k}$,
showing that the nucleation-controlled case is obtained by replacing
$l_{2k}$ with $L.$ 

It was proposed \citep{Hirth,Seeger-1981} to capture the system size
dependence by multiplying $v_{2k}$ by the interpolating function
$L/(L+l_{2k})$ with $l_{2k}$ calculated for an infinitely large
system. It is easy to show that 
\begin{equation}
l_{2k}=\left(\dfrac{2v}{aJ}\right)^{1/2}=a\left(\dfrac{P_{10}+2(P_{12}-P_{21})}{P_{01}}\right)^{1/2}.\label{eq:GB-vel-scaling}
\end{equation}
Comparison with the KMC simulations shows that this interpolating
function overcorrects the model by under-predicting the GB velocity
(not shown in Fig.~\ref{fig:Velocity-force-relations}). Note that
the choice of the interpolating function is arbitrary as long as it
gives the correct values of 1 and $L/l_{2k}$ in the limits of $l_{2k}\ll L$
(infinitely large system) and $l_{2k}\gg L$ (small system), respectively.
For example, we find that the function $L/(L^{2}+l_{2k}^{2})^{1/2}$
provides much better agreement with the simulations of high GB energies
(Fig.~\ref{fig:kink-mechanism}). The respective GB velocity is 
\begin{equation}
v_{2k}=2v\left(\dfrac{P_{01}}{\left[P_{10}+2(P_{12}-P_{21})\right]\left(1+l_{2k}^{2}/L^{2}\right)}\right)^{1/2}\label{eq:GB_vel-1-1}
\end{equation}
with $l_{2k}$ given by Eq.(\ref{eq:GB-vel-scaling}).

The demonstrated agreement between the kink pair model and the simulation
results corroborates our explanation that the peculiar S-shape of
the velocity-force relations in Fig.~\ref{fig:Velocity-force-relations}
is caused by a transition of the GB migration mechanism from (1) kink
pair nucleation and growth under a small force to (2) driven random
walk of a rough GB structure under a larger force.

\subsection{GB dynamics with pinning}

We now consider a GB subject to the pinning effect. It is convenient
to discuss the impact of pinning in terms of the normalized solute
diffusivity $D/D_{0}$ rather than the normalized pinning time $\tau_{p}/\tau_{0}$
as in section \ref{GB-properties}. The relation between the two is
$D/D_{0}=\tau_{0}/\tau_{p}$ \citep{Mishin_2023_RW_part_I}. In the
simulations presented below, $D/D_{0}$ was varied while the pinning
strength was fixed at $\alpha=1.5$.

As expected, we find that the pinning always reduces the GB velocity
under a given driving force relative to the unpinned case (Fig.~\ref{fig:Velocity-force-relations-drag}),
which is a manifestation of the solute drag effect. As also expected,
faster solute diffusion (larger $D/D_{0}$) causes stronger retardation
of the GB motion. This is understandable because a faster solute can
keep up with the moving GB and slow it down. The S-shape of the velocity-force
curves becomes more pronounced, especially for high-energy GBs (Fig.~\ref{fig:Velocity-force-relations-drag}(b)).
As discussed in the previous subsection, the S-shape is caused by
the dynamic roughening effect. Since the pinning reduces the GB velocity,
it partially suppresses the dynamic roughening. The slow kink pair
migration mechanism continues to operate until larger forces, causing
the nearly flat portion of the curves in the low-velocity regime.

The solute drag force is defined as the difference between the force
required to drive the GB in the presence of pinning and the force
to drive an unpinned GB with the same velocity \citep{Mishin_2023_RW_part_I}.
In Fig.~\ref{fig:Velocity-force-relations-drag}, the drag force
is the horizontal distance between the pinned curves and the unpinned
curve corresponding to $D/D_{0}=0$. The velocity dependence of the
normalized drag force $\varphi_{d}$ is shown in Fig.~\ref{fig:Velocity-vs-force}
for several values of $D/D_{0}$. For the low-energy GB ($\sigma=0.1$),
the drag-velocity curves look qualitatively similar to those in the
1D model \citep{Mishin_2023_RW_part_I}, except that the magnitude
of the drag force is systematically higher. As long as the solute
diffusivity is not too high, the drag force reaches a maximum at a
critical velocity, as predicted by the classical solute drag models
\citep{Cahn-1962,Lucke-Stuwe-1963,Lucke:1971aa} and confirmed in
the 1D version of the present model \citep{Mishin_2023_RW_part_I}.
Recall that on the left of the maximum, the GB drags the segregation
atmosphere, while on the right of the maximum, it breaks away from
it. 

As in the 1D case \citep{Mishin_2023_RW_part_I}, some results of
the 2D simulations deviate from predictions of the classical models.
This includes Cahn's \citep{Cahn-1962} prediction that the maximum
drag force is independent of the solute diffusivity. Figure \ref{fig:Velocity-vs-force}
shows that the maximum value of $\varphi_{d}$ strongly depends on
the solute diffusivity. Fast solute diffusion amplifies the drag by
increasing the height of the maximum and shifting it towards larger
velocities. When $D/D_{0}$ is large enough, the maximum smooths out.
The breakaway regime disappears, and the drag force becomes a monotonically
increasing function of GB velocity. The atmosphere remains permanently
attached to the GB and evolves continuously from heavy when the GB
moves slowly to light when it moves fast.

An important effect revealed by the present simulations and not captured
by 1D models is the impact GB roughness on the solute drag. As shown
in Fig.~\ref{fig:Velocity-vs-force}, the drag-velocity curves for
the high-energy GB ($\sigma=4$) have a significantly different shape
than those for the low-energy boundary ($\sigma=0.1$). Recall that
the former is smooth in the stationary state while the latter is rough.
The difference between the two cases increases as the solute diffusivity
decreases. For the high-energy GB, the low-velocity portions of curves
become nearly vertical and exhibit a behavior akin to a threshold
effect. Namely, the GB velocity remains extremely low until the drag
force reaches a critical (threshold) level. At the critical force,
the GB abruptly accelerates, producing the nearly horizontal portion
on the curves. This transition is continuous but very sharp. It is
further illustrated in Fig.~\ref{fig:Solute-drag-2D}, where it is
especially pronounced for the high-energy GB at $D/D_{0}=0.05$. Observe
that the transition occurs on the low-velocity side of the drag-force
maximum where the atmosphere is attached to the GB. Note also that,
on the high-velocity side of the maximum, the high- and low-energy
curves converge to each other, demonstrating similar dynamics.

\begin{figure}
\begin{centering}
\includegraphics[width=0.6 \textwidth]{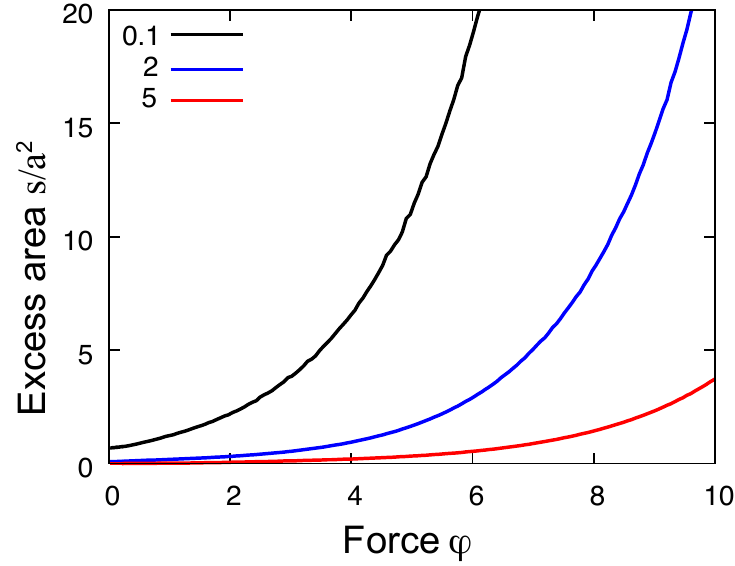}
\par\end{centering}
\caption{Excess GB area $s$ (normalized by $a^{2}$) as a function of driving
force at the temperature of $\theta=0.2$ for three values of the
GB energy $\sigma$. At each value of the force, $s$ was obtained
by averaging over a long time in the steady-state regime. The system
size is $N=512$ and the GB is not subject to pinning. \label{fig:Excess-interface-area}}
\end{figure}

\begin{figure}
\includegraphics[width=0.55\textwidth]{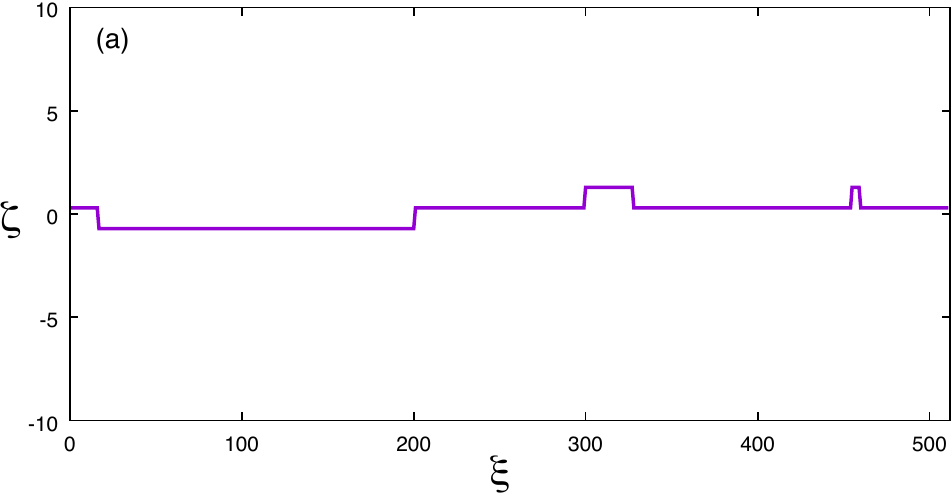}

\medskip{}

\includegraphics[width=0.55\textwidth]{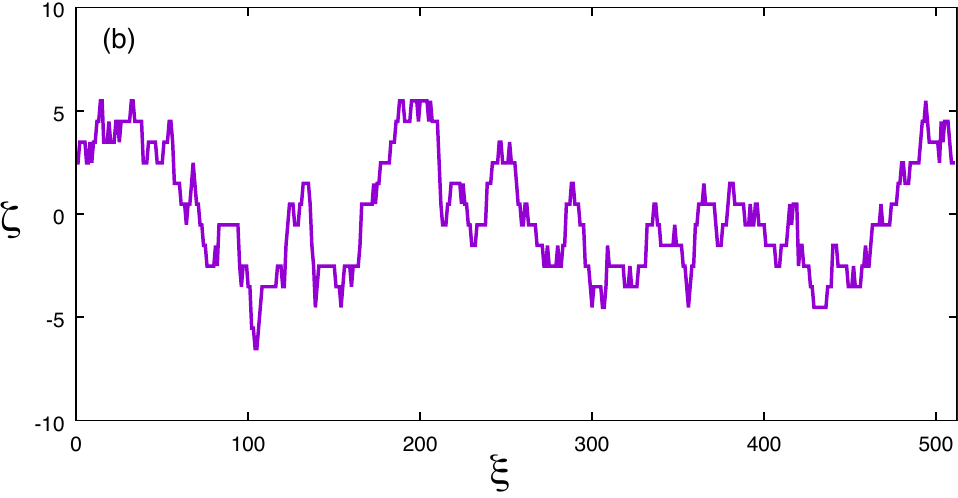}

\medskip{}

\includegraphics[width=0.55\textwidth]{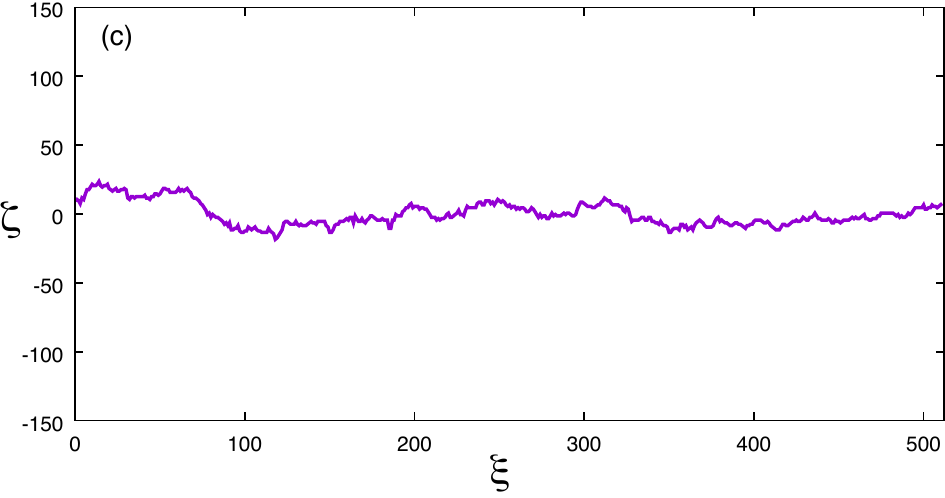}

\medskip{}

\includegraphics[width=0.55\textwidth]{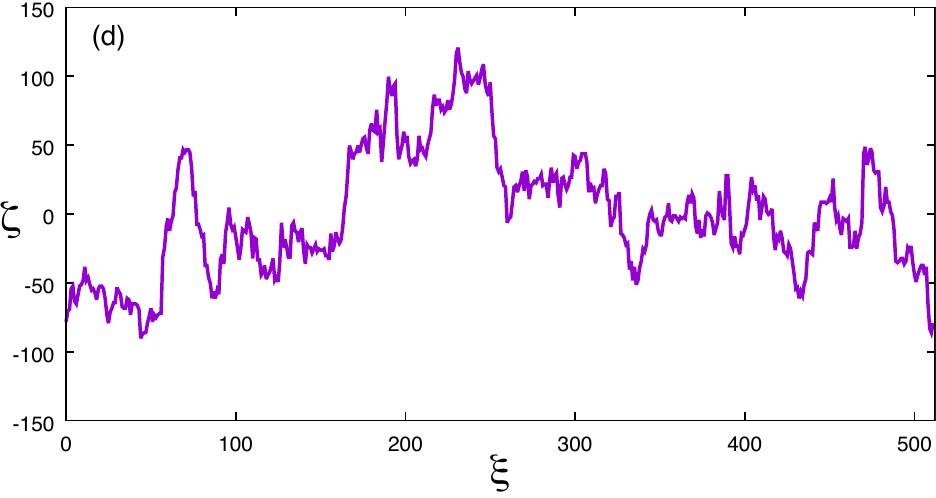}

\caption{Demonstration of dynamic GB roughening at the temperature of $\theta=0.2$
without pinning. The initially smooth GB ($\sigma=5$) remains smooth
under a force of $\varphi=0.2$ (a) but becomes rough at $\varphi=4$
(b). The initially rough boundary ($\sigma=0.1$) remains equally
rough at $\varphi=0.2$ (c) but becomes much rougher at $\varphi=4$
(d). The system size is $N=512$. Note the difference in the scale
of the $\varsigma$-axes. \label{fig:Dynamic-roughness}}
\end{figure}

\begin{figure}
\begin{centering}
\includegraphics[width=0.7 \textwidth]{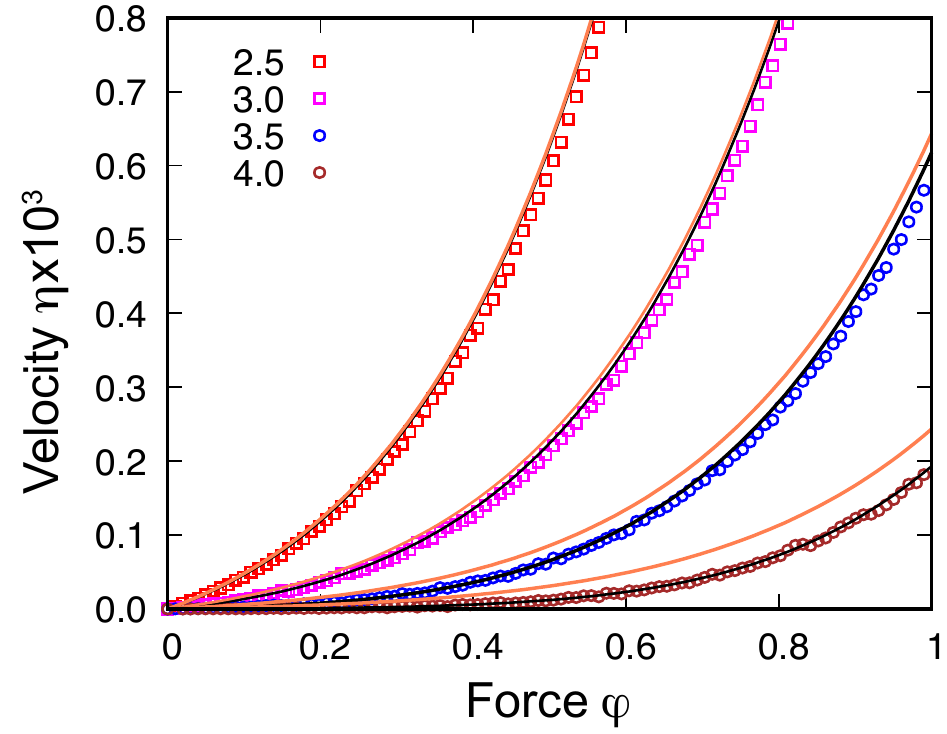}
\par\end{centering}
\caption{Zoom into the low-velocity portions of the velocity-force curves shown
in Fig.~\ref{fig:Velocity-force-relations}. The GB energies $\sigma$
are indicated in the key. The points represent KMC simulation runs.
The orange curves are predicted by Eq.(\ref{eq:GB_vel-1}) based on
the kink pair GB migration model neglecting the system size effect.
The black curves are predicted by Eq.(\ref{eq:GB_vel-1-1}) including
the system size correction. \label{fig:kink-mechanism}}
\end{figure}

\begin{figure}
\begin{centering}
(a) \includegraphics[width=0.65\textwidth]{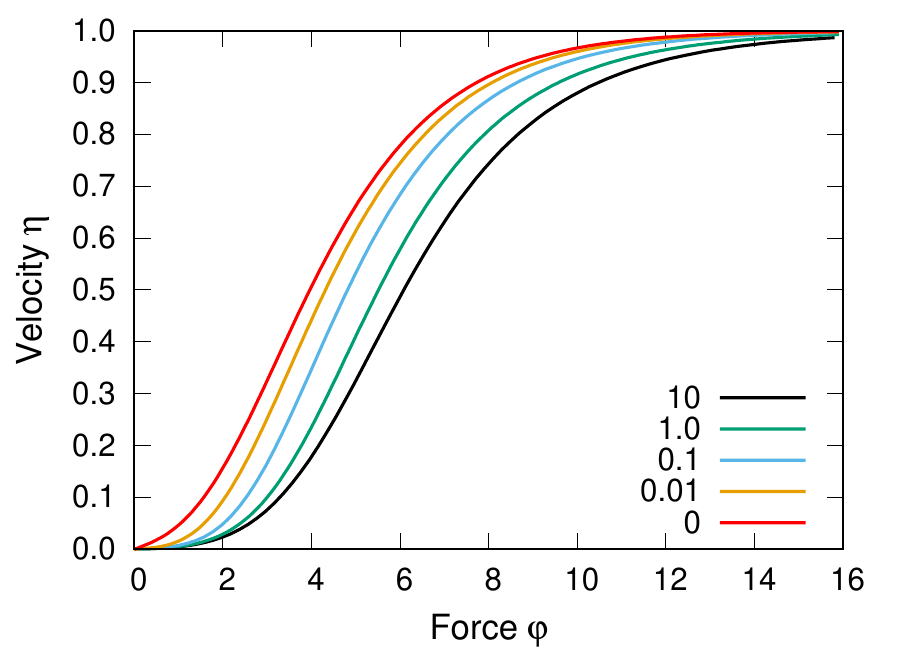}
\par\end{centering}
\begin{centering}
\bigskip{}
\par\end{centering}
\begin{centering}
(b) \includegraphics[width=0.65\textwidth]{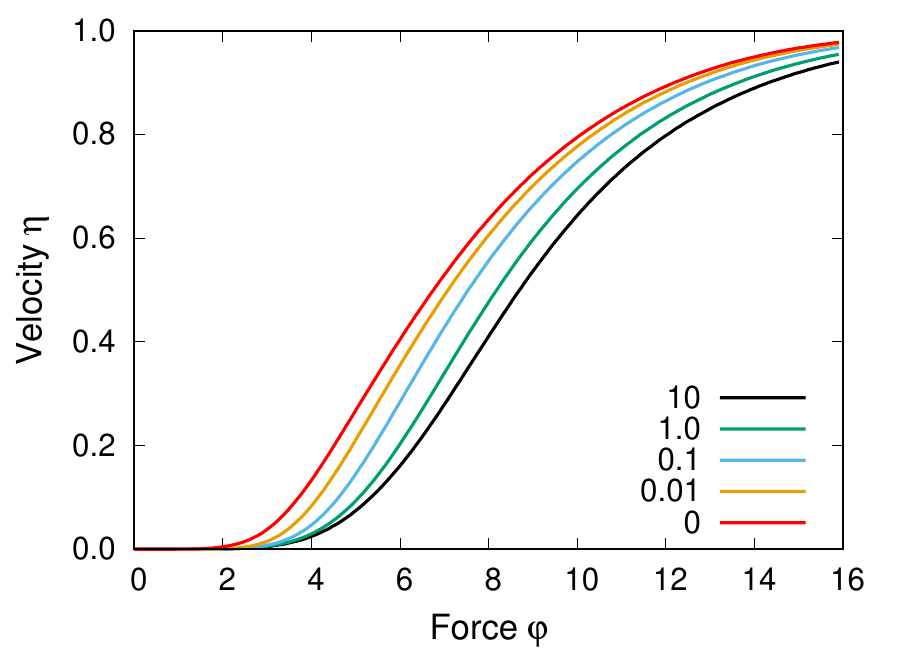}
\par\end{centering}
\begin{centering}
\caption{Velocity-force relations for the 2D GB with $N=512$ nodes at the
temperature of $\theta=0.2$ for several values of the normalized
solute diffusivity $D/D_{0}$ indicated in the key. The curve for
$D/D_{0}=0$ was obtained by simulations without pinning. (a) $\sigma=0.1$;
(b) $\sigma=4.0$. \label{fig:Velocity-force-relations-drag}}
\par\end{centering}
\end{figure}

\begin{figure}
\begin{centering}
(a) \includegraphics[width=0.5 \textwidth]{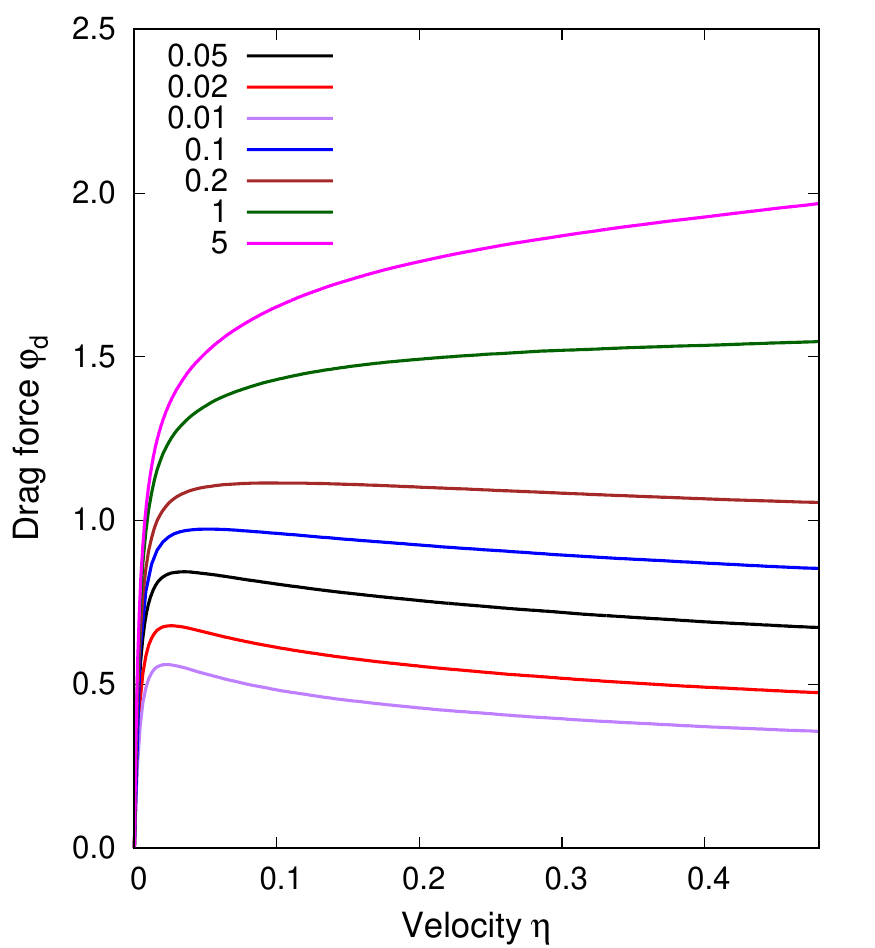}
\par\end{centering}
\begin{centering}
\bigskip{}
\par\end{centering}
\begin{centering}
(b) \includegraphics[width=0.5\textwidth]{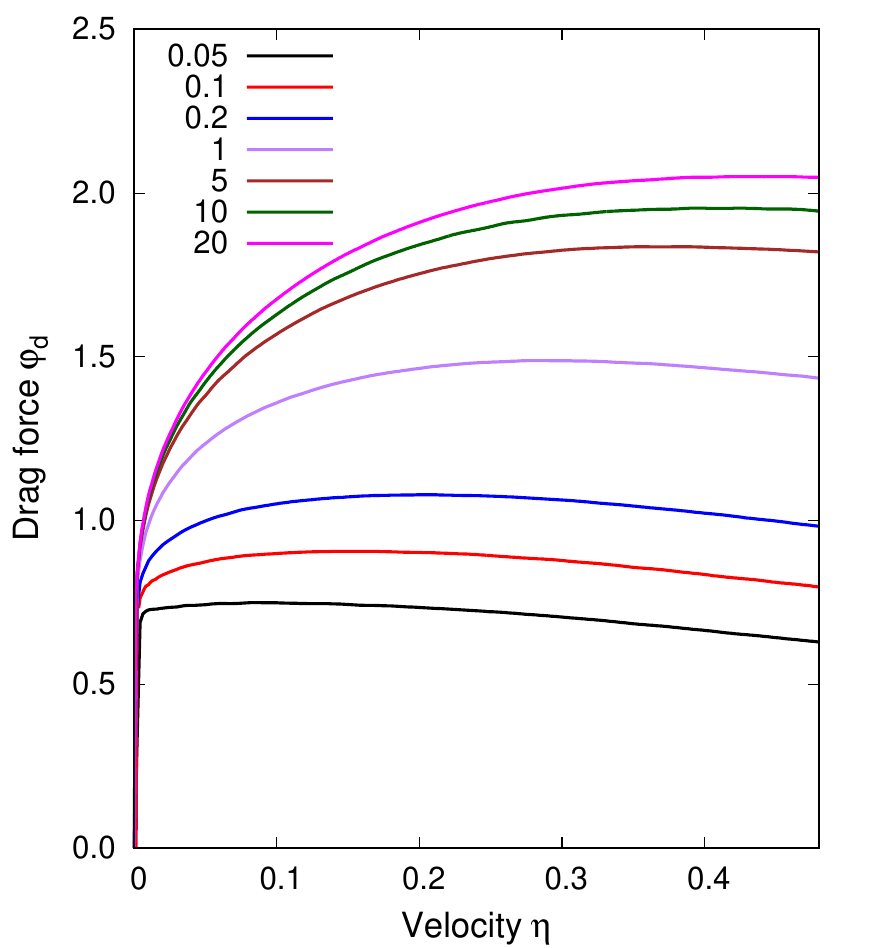}
\par\end{centering}
\centering{}\caption{Solute drag force $\varphi_{d}$ as a function of velocity $\eta$
for the 2D GB with $N=512$ nodes at the temperature of $\theta=0.2$
for several values of the normalized solute diffusivity $D/D_{0}$
indicated in the key. (a) $\sigma=0.1$; (b) $\sigma=4.0$. \label{fig:Velocity-vs-force}}
\end{figure}

\begin{figure}
\begin{centering}
\includegraphics[width=0.6\textwidth]{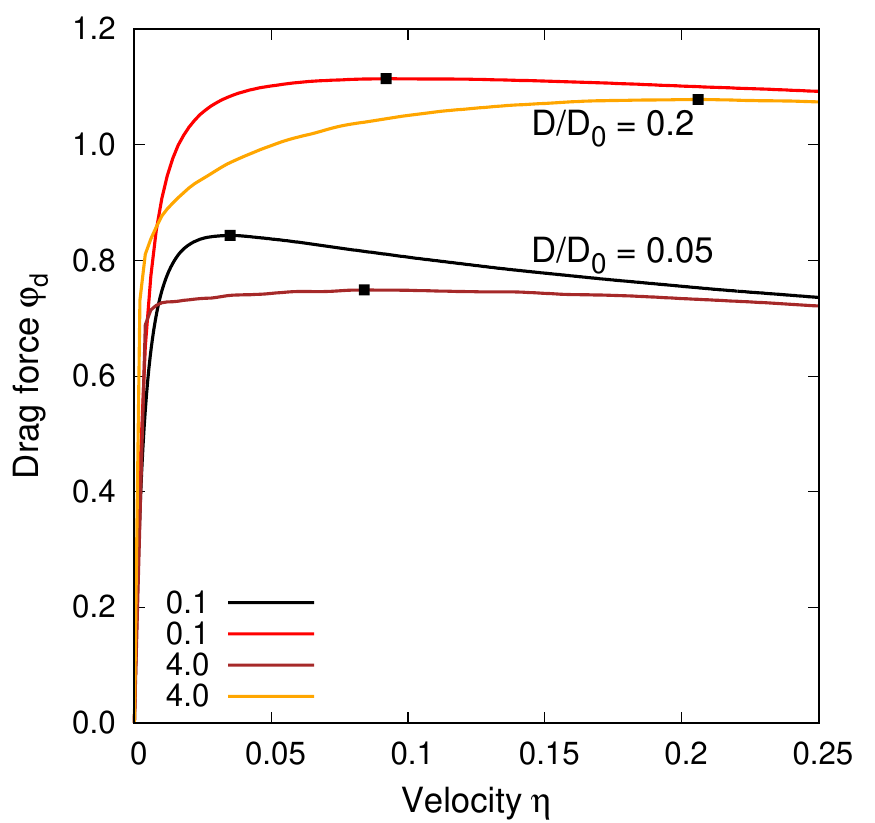}
\par\end{centering}
\caption{Solute drag force $\varphi_{d}$ as a function of velocity $\eta$
for the 2D GB with $N=512$ nodes at the temperature of $\theta=0.2$
for two values of the normalized solute diffusivity $D/D_{0}$. The
GB energies $\sigma$ are indicated in the key. The points mark the
maxima of $\varphi_{d}$.\label{fig:Solute-drag-2D}}
\end{figure}

\begin{figure}
\begin{centering}
(a) \includegraphics[width=0.55\textwidth]{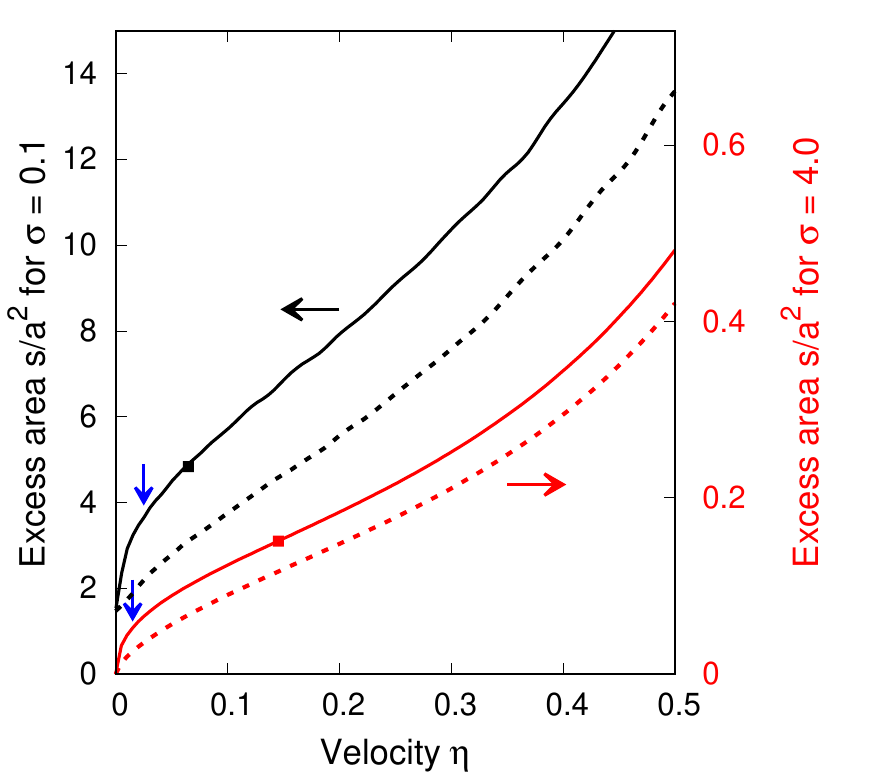}
\par\end{centering}
\bigskip{}

\begin{centering}
(b) \includegraphics[width=0.56\textwidth]{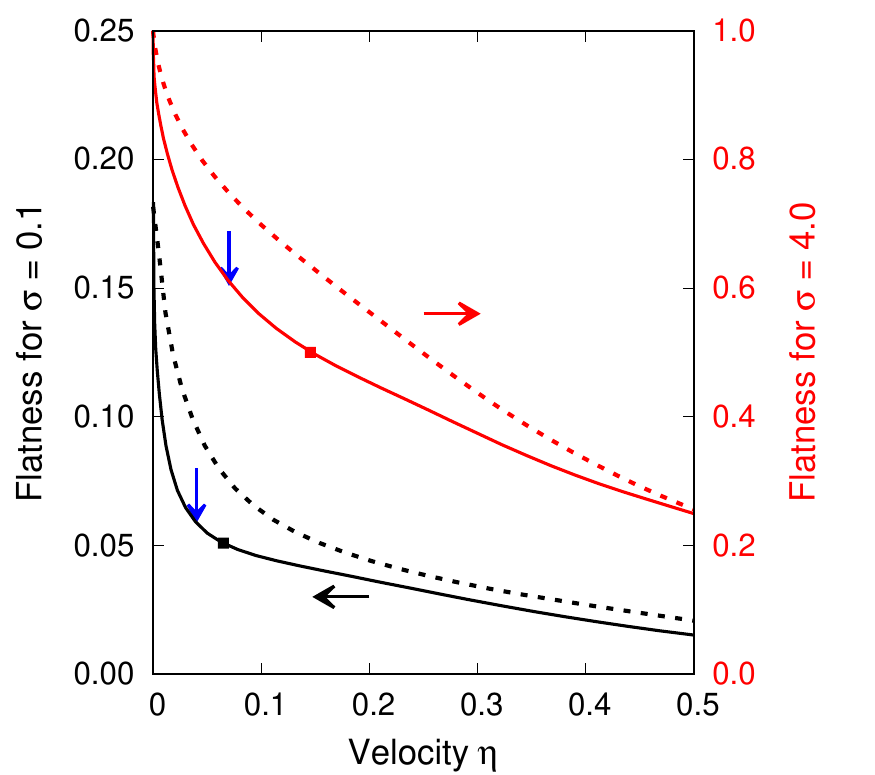}
\par\end{centering}
\caption{(a) Excess GB area $s/a^{2}$ and (b) GB flatness parameter $f$ as
a function GB velocity $\eta$ for the GB energies $\sigma=0.1$ (black
curves) and $\sigma=4.0$ (red curves). $s$, $f$ and $\eta$ were
obtained by averaging over a long period of time after the GB motion
reaches a steady state. The solid curves represent an alloy with the
normalized solute diffusivity $D/D_{0}=0.1$. The dashed curves were
obtained by solute-free simulations. The points mark the velocities
at which the solute drag force reaches a maximum. The vertical blue
arrows indicate approximate locations of the dynamic roughening transition.
The temperature is $\theta=0.2$ and the system size is $N=512$.
\label{fig:Excess-interface-area-drag}}
\end{figure}

Analysis shows that the threshold behavior of the solute drag is caused
by the dynamic roughening effect. The latter was previously demonstrated
for unpinned GBs (see Figs.~\ref{fig:Excess-interface-area} and
\ref{fig:Dynamic-roughness}). It was shown that a smooth GB becomes
rough when driven by an external force. This effect is reproduced
in Fig.~\ref{fig:Excess-interface-area-drag}, where we use the excess
GB area and the GB flatness parameter $f$ as measures of roughness.
This time we include the simulation results obtained in the presence
of pinning. The plots show that the excess area increases and the
flatness decreases with the GB velocity in all cases, which is a manifestation
of dynamic roughening. We also observe that pinning increases the
GB roughness relative to the unpinned GB moving with the same velocity. 

Furthermore, in the presence of pinning, the curves in Fig.~\ref{fig:Excess-interface-area-drag}
tend to develop a threshold behavior in the low-velocity limit. The
nearly vertical portion of the curves indicates that the GB resists
the motion. As the driving force increases, so does the GB roughness
due to the reduced barriers for kink pair nucleation. The GB remains
nearly pinned in place until it develops a sufficient degree of roughness.
Once a high enough level of roughness is reached, the GB motion accelerates,
which in turn causes further roughening. Eventually, the GB enters
a kinetic regime in which its dynamic roughness increases gradually
with the velocity. 

Although the dynamic roughening transition is continuous, it seems
reasonable to associate it with the point of maximum curvature on
the roughness-velocity curves. In Fig.~\ref{fig:Excess-interface-area-drag},
such points are marked by vertical arrows. We emphasize that the dynamic
roughening transition differs from the previously discussed static
roughening transition (which is thermodynamic in nature) both conceptually
and in the degree of roughness reached at the transition point. Dynamic
roughening also occurs without pinning when it is more diffuse (spread
over a wider interval of velocities). The pinning shifts the transition
towards smaller velocities and makes it much sharper. This causes
the threshold behavior of the solute drag seen in Figs.~\ref{fig:Excess-interface-area}
and \ref{fig:Dynamic-roughness}. Thus, the dynamic roughening transition
in GBs can strongly impact the solute drag dynamics, especially for
high-energy GBs that are smooth under stationary conditions.

\section{Discussion and conclusions}

We have developed a stochastic model of solute-GB interactions aiming
to better understand the solute drag phenomenon. The model is very
simple and contains only a small number of parameters but still captures
the main physics of solute interactions with both stationary and moving
GBs. The model describes the kinetic competition between GB migration
and solute diffusion, which is the key mechanism of the solute drag.
The solute diffusion is included in the model indirectly through the
square root time dependence of the GB jump barriers. The GB dynamics
is represented more accurately than in the linear approximation commonly
employed in the modeling of GB migration. 

Since the model is stochastic, its numerical solution requires KMC
simulations. At variance to the traditional KMC simulations, the present
random walk algorithm does not implement a Markov chain. As discussed
in Part I \citep{Mishin_2023_RW_part_I}, the transitions between
the GB states are memoryless but the residence time does not follow
the exponential distribution, making the random walk a semi-Markov
process \citep{Chari:1994aa,Maes:2009aa,Yu:2010aa}. As a consequence,
the steady-state occupation probabilities of the GB states do not
follow the Boltzmann distribution. This does not contradict the equilibrium
statistical mechanics because the GB never reaches the true equilibrium.
It is not only coupled to a thermostat but also interacts with a reservoir
of solute atoms in a rate-dependent manner. The impact of this interaction
on the steady-state occupation probabilities is strongest when the
pinning time scale $\tau_{p}$ is close to the time scale $t_{0}$
of equilibrium thermal fluctuations, a situation that we call active
pinning. We expect that non-Markovian behavior is a common feature
of all systems exhibiting diffusion-controlled interactions with segregating
solutes.

The 2D version of the model represents the GB as a solid-on-solid
(SOS) interface. This choice is not the only possible option: some
of the previous simulations of GB migration utilized the Ising model
\citep{Mendelev02a,Mendelev:2001aa,Mendelev:2001wk,Mendelev:2002wv}.
The latter has certain advantages as well as drawbacks relative to
SOS models. One advantage is that the results can be mapped more easily
onto other applications of the Ising model in many areas of physics.
This can help interpret the results and borrow computer algorithms
and experience from other fields. Furthermore, the Ising model can
represent a smooth or faceted shape of an entire 2D or 3D grain. On
the other hand, the Ising model can create GB protrusions with overhands
and, in some cases, isolated inclusions (``bubbles'') of one grain
inside the other. Such features do not reflect the typical morphologies
of moving GBs. SOS models describe the GB shape by a \emph{single-valued}
function $z(x,y)$ avoiding overhangs and inclusions. This description
is appropriate for individual moderately curved portions of a GB,
but an entire grain cannot be represented. It should also be noted
that SOS models are computationally faster because the KMC attempts
are only made at the interface, whereas in the Ising models the ``spin
flip'' attempts have to be made across the entire system. The computational
efficiency of SOS models provides access to larger systems and enables
a broader exploration of the parameter space. 

The model reproduces a roughening transition in stationary GBs as
temperature increases. This is a continuous transition that can be
identified with the peak of the GB heat capacity. Active pinning reduces
the roughening transition temperature and makes the transition sharper.
The model also predicts dynamic roughening, a process in which a smooth
GB becomes rough as the migration velocity increases at a fixed temperature.
This is another continuous and fully reversible transition: the GB
returns to the smooth state if it comes to rest. Without pinning,
the dynamic roughening transition is spread over a broad velocity
range. The pinning shifts this transition towards lower velocities
and makes it significantly sharper. 

The mechanism of dynamic roughening has been studied in great detail.
In a stationary state or when the velocity is low, GB migration occurs
by the kink pair mechanism, which can only sustain slow motion. The
driving force reduces the barrier for kink pair nucleation in the
forward direction, boosting the population of non-equilibrium kinks
and increasing the GB mobility. In section \ref{subsec:GB-dynamics},
we proposed an analytical model of this process that agrees well with
the simulation results. The growing kink concentration and its spatial
variations eventually cause capillary waves. The GB structure becomes
rough and the GB migration mechanism changes from kink-mediated to
a random walk of the average GB plane. The dynamic roughening transition
is responsible for the threshold behavior in GB dynamics, in which
the GB moves very slowly until the driving force reaches a critical
level at which the motion sharply accelerates. This threshold effect
is especially strong for high-energy GBs. The solute pinning sharpens
this effect and increases the force required to ``unlock'' the GB
mobility. Somewhat similar shapes of drag-velocity curves were observed
in prior KMC simulations using different methodologies \citep{Mendelev02a,Mendelev:2001aa,Mendelev:2001wk,Mendelev:2002wv,Wicaksono:2013aa}.

To put our results in perspective with the literature, dynamic roughening
of open surfaces is known in the field of crystal growth \citep{Balibar:1990aa,Tartaglino:2000aa}.
For GBs in materials, a threshold behavior similar to the one found
in this paper was observed experimentally; see examples in Figures
2 and 5 in Ref.~\citep{KANG:2016aa}, where the GB structures below
and above the threshold were characterized as faceted (smooth, atomically
ordered) and rough, respectively. The impact of roughening on GB mobility
was studied by molecular dynamics (MD) simulations \citep{Olmsted07a}.
The GB mobility in Ni was found to be much higher above the roughening
transition than below. In another MD study \citep{Marian:2004aa},
screw dislocations in body-centered cubic metals underwent a dynamic
transition from kink pair mediated glide at small strain rates to
jerky motion of a rough dislocation line at high strain rates. Of
course, dislocations present a different case for many reasons, including
the prominent role of the elastic strain field and the kink-kink interactions.
Nevertheless, this transition is likely another manifestation of the
dynamic roughening phenomenon discussed here. 

Observing static or dynamic roughening in simulations or experiment
requires a particular combination of parameters, such as the GB energy,
GB mobility, and (in alloys) solute segregation and solute diffusivity.
Some GBs can remain smooth all the way to the melting point, while
others premelt before they could undergo a roughening transition.
GB faceting is another transition that can interplay with roughening
but remains beyond the scope of this paper.

Both the 1D and 2D versions of the model reveal effects that were
not in the classical models \citep{Cahn-1962,Lucke-Stuwe-1963,Lucke:1971aa}.
One of them is the increase of the maximum drag force with the solute
diffusivity. Further, the classical models do not capture the GB roughening
and its impact on the GB migration mechanisms. Another difference
is related to the breakaway branch of the drag-velocity relation.
Cahn \citep{Cahn-1962} considered this branch unstable with respect
to velocity variations. He reasoned that, if the GB momentarily moves
faster, it will lose some of the segregation atmosphere, which will
allow it to move even faster. Recall that Cahn's model treats the
GB as a rigid plane. In the 2D version of our model, velocity fluctuations
do occur locally, but they are suppressed by the interface tension
and do not develop into a morphological instability. GB motion in
the breakaway regime remains perfectly stable. The GB shape fluctuations
indeed grow with the GB velocity, as seen on the roughness-velocity
plots (Fig.~\ref{fig:Excess-interface-area-drag}). However, this
increase is gradual and occurs at about the same rate as without pinning. 

It should be recognized that the discreteness of the GB displacements
imposed by the underlying $a\times a$ grid is a critical ingredient
of the model capturing the existence of the GB structural units. It
is due to this discreteness that the GB can be smooth or rough and
can migrate by the kink pair mechanism. Details of the kink structure
and energy may depend on the grid structure and symmetry. However,
without the grid, there would be no kinks and no roughening transition.

Being very simple, the proposed model is not intended for accurate
quantitative predictions for a particular material. Nevertheless,
the numerical values of the parameters chosen for this work are quite
realistic. For example, it was found above that the roughening transition
temperature $T_{r}$ satisfies the condition $\theta_{r}/\sigma=k_{B}T_{r}/\gamma a^{2}\approx0.2$.
Taking Cu as an example, we can estimate $a\approx0.3$ nm (between
the first and second neighbors). GB energies in Cu vary widely from
$\sim0.2$ J\,m$^{-2}$ for low-angle GBs to $\sim1$ J\,m$^{-2}$
for some of the high-angle, high-energy GBs \citep{Cahn06b,Frolov2012b,Frolov2013,Hickman:2016aa,Hickman__2017a,Koju:2021aa}.
Taking $\gamma=0.6$ J\,m$^{-2}$ as a representative value, we obtain
$\gamma a^{2}\approx0.48$ eV and thus $T_{r}\approx780~K=0.57T_{m}$.
This is a meaningful temperature at which high-angle GBs in Cu are
just beginning to develop structural disorder \citep{Cahn06b,Frolov2012b,Frolov2013,Hickman:2016aa,Frolov:2015aa}.
The GB displacement barrier $E_{0}$ can be associated with the activation
energy of GB migration. The latter also varies widely, depending on
the GB crystallography, temperature, and the presence of extrinsic
defects \citep{Race:2014aa,Race:2015aa,Hadian:2016aa,Koju:2021aa,Homer:2022aa}.
Analysis of literature data for Cu and other metals (rescaled by the
melting temperature) shows that $E_{0}$ for Cu lies roughly between
0.2 and 1.5 eV. GB segregation is known to strongly increase the migration
energy. For example, adding only 1 at.\%Ag increases the migration
barrier of the $\Sigma$17 {[}001{]} tilt GB in Cu from 0.47 eV to
1.5 eV \citep{Koju:2021aa}. Assuming that the GB segregation is close
to saturation, the respective pinning factor $\alpha$ varies between
1 and 3.2. Most of the simulations in this work used $\alpha=1.5$,
which is well within the range of physically meaningful values.

In conclusion, we presented a model that provides useful insights
into GB interactions with solutes in general and the impact of GB
roughening on the GB dynamics, GB migration mechanisms, and the solute
drag. It is hoped that this model can inform future modeling studies
targeting specific materials and possibly motivate new experiments.
It would also be interesting to extend the model to 3D, in which case
the GB roughening becomes a real phase transformation.

\bigskip{}

\noindent \textbf{Acknowledgements} 

This research was supported by the National Science Foundation, Division
of Materials Research, under Award no. 2103431.

\section*{Appendix A: The capillary fluctuation theory}

In this appendix, we briefly summarize the main results of the capillary
fluctuation theory for 1D interfaces.

The starting point is the free energy of a $L\times a$ ($L\gg a$)
interface lying in the $(x,y)$ plane \citep{Gelfand:1990vz,Lapujoulade:1994uw,Saito:1996vo}:
\begin{equation}
\mathcal{F}=f_{0}La+\dfrac{1}{2}\Lambda a\int_{0}^{L}\left(\dfrac{\partial w}{\partial x}\right)^{2}dx,\label{eq:3.1}
\end{equation}
where $f_{0}$ is the free energy per unit area of a planar interface,
$\Lambda$ is the interface stiffness defined by Eq.(\ref{eq:3.2-1})
in the main text, and $w(x)=z(x)-z_{*}$ is the local shape deviation
from the planar geometry. Suppose the interface fluctuation profile
$w(x)$ is periodic with the period $L$ and is represented by $N$
points $\{x_{j},w_{j}\}$, where $x_{j}=ja$, $j=0,1,...,N-1$, and
$w_{N}=w_{0}$. This profile can be approximated by the discrete Fourier
series 
\begin{equation}
w(x)=\sum_{n=-(N-1)/2}^{(N-1)/2}\hat{w}_{n}e^{-ik_{n}x}.\label{eq:3.3}
\end{equation}
with the wave numbers $k_{n}=(2\pi/L)n$. (We assumed for simplicity
that $N$ is odd.) The complex Fourier amplitudes $\hat{w}_{n}$ satisfy
the relation $\hat{w}_{-n}=\hat{w}_{n}^{*}$ with $\hat{w}_{0}=0$.
Inserting the Fourier expansion (\ref{eq:3.3}) in Eq.(\ref{eq:3.1})
and using the orthogonality of the basis functions, we obtain
\begin{equation}
\mathcal{F}=f_{0}La+\dfrac{1}{2}\Lambda La\sum_{n=-(N-1)/2}^{(N-1)/2}\left|\hat{w}_{n}\right|^{2}k_{n}^{2}\label{eq:3.5}
\end{equation}
The $(N-1)$ nonzero terms in this expansion represent decoupled vibrational
modes. Since each term is quadratic in the fluctuation amplitude $\left|\hat{w}_{n}\right|$,
the canonical ensemble-averaged square fluctuation is given by \citep{Mishin:2015ab}
\begin{equation}
\overline{\left|\hat{w}_{n}\right|^{2}}=\dfrac{k_{B}T}{\Lambda Lak_{n}^{2}},\quad n=-\dfrac{N-1}{2},...,\dfrac{N-1}{2}.\label{eq:3.8}
\end{equation}
Inserting this fluctuation spectrum into Parseval's theorem
\begin{equation}
\dfrac{1}{N}\sum_{n=-(N-1)/2}^{(N-1)/2}w_{n}^{2}=\sum_{n=-(N-1)/2}^{(N-1)/2}\left|\hat{w}_{n}\right|^{2},\label{eq:3.9}
\end{equation}
we obtain the mean squared interface width
\begin{equation}
\overline{w^{2}}=\sum_{n=-(N-1)/2}^{(N-1)/2}\dfrac{k_{B}T}{\Lambda Lak_{n}^{2}}=2\dfrac{k_{B}TL}{4\pi^{2}\Lambda a}\left(\sum_{n=1}^{(N-1)/2}\dfrac{1}{n^{2}}\right).\label{eq:3.10}
\end{equation}
Note that 
\begin{equation}
\sum_{n=1}^{\infty}\dfrac{1}{n^{2}}=\dfrac{\pi^{2}}{6}.\label{eq:3.13}
\end{equation}
In the thermodynamic limit ($N\rightarrow\infty$), Eq.(\ref{eq:3.10})
converges to Eq.(\ref{eq:3.12-1}) of the main text. Accordingly,
the interface free energy per unit area becomes
\begin{equation}
\dfrac{\mathcal{F}}{La}=f_{0}+\dfrac{k_{B}T}{a^{2}},\label{eq:3.14}
\end{equation}
where the second term represents the capillary-wave contribution.

\section*{Appendix B: Survival probability a kink pair }

In this Appendix we derive Eq.(\ref{eq:Escape_prob}) of the main
text for the survival probability $P_{s}$ of the triangular bump,
which represents a kink pair nucleus on a planar GB driven by an applied
force. We will first calculate the probability $P_{c}$ that the triangular
bump disappears creating a planar GB. Then $P_{s}=1-P_{c}$.

\begin{figure}
\begin{centering}
\includegraphics[width=0.9\textwidth]{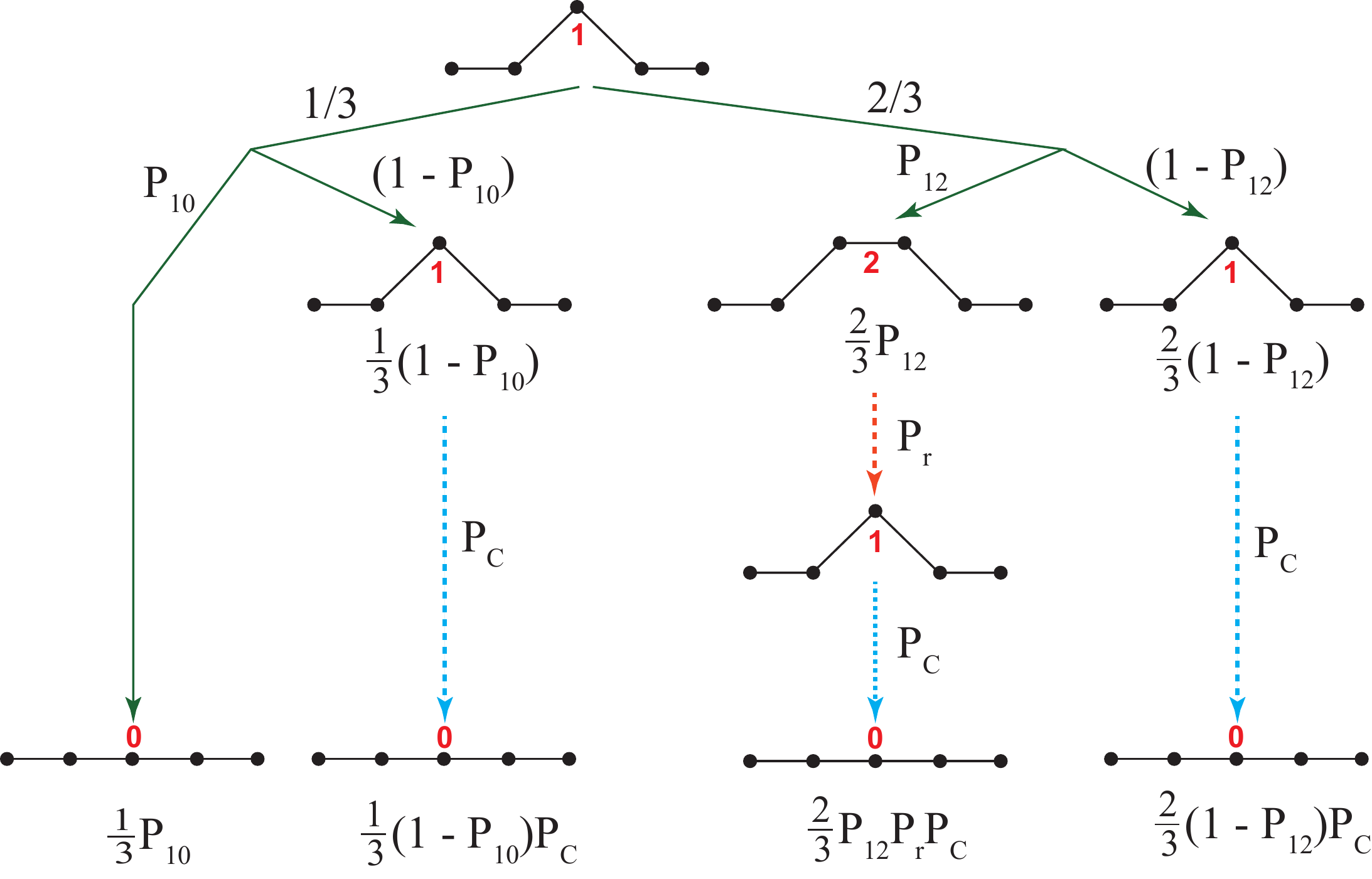}
\par\end{centering}
\caption{Event diagram for calculating the survival probability of a kink pair
nucleus (triangular bump) on a planar GB moving under an applied force
pointing upward. The formulas on the diagram represent the probabilities
of different states of the kink pair and transitions between them.
\label{fig:Event-diagram}}

\end{figure}

The calculation is explained on the event diagram in Fig.~\ref{fig:Event-diagram}.
The following notation is used. A planar GB, a triangular bump, and
a two-node kink pair are referred to as states 0, 1 and 2, respectively,
according to the number of nodes above ground level. These states
are labeled by red numerals. The formulas represent the probabilities
of different states and transitions (jumps) between the states. The
applied force is pointing upward, and the initial state of the system
is a triangular bump (state 1) shown on top of the diagram. Only nodes
belonging to the kinks are allowed to jump and only in a manner that
preserves the single-layer height of all kinks above the ground.

It is convenient to describe the system evolution as occurring during
a KMC simulation. At the first KMC step, one of the three nodes of
the triangular bump is selected at random. The subsequent events are
represented by the solid green arrows (Fig.~\ref{fig:Event-diagram}).
The tip node is selected with a probability 1/3. Once selected, the
tip node can jump down with the probability $P_{10}$ or remain intact
with the probability $(1-P_{10})$. If the jump attempt is successful,
the triangular bump disappears. The probability of this outcome is
$(1/3)P_{10}$. If the attempt fails, the triangular bump survives
the first KMC step with the probability $(1/3)(1-P_{10})$. In this
case, the bump can still collapse during the subsequent evolution
with the (still unknown) probability $P_{c}$. This collapse can happen
after a chain of jumps shown on the diagram by the dashed blue arrow.
The collapse probability of the triangular bump along this route is
$(1/3)(1-P_{10})P_{c}$. 

Returning to the first KMC step, there is a 2/3 chance that one of
the two base nodes of the triangular bump is selected. This node will
then attempt to jump upward and create a two-node kink pair (state
2). The success probability of this jump is $P_{12}$, so a two-node
kink pair can form with the probability of $(2/3)P_{12}$. Once formed,
this kink pair can grow further or transform back into a triangular
bump. The back transformation can happen immediately (probability
$P_{21}$) or after some period of growth represented by the dashed
red arrow. Let $P_{r}$ be the probability of returning into the triangular
bump. The latter can then collapse into a planar GB. The collapse
probability along this route is thus $(2/3)P_{12}P_{r}P_{c}$. Finally,
if the selected base node cannot make a successful jump, the triangular
bump remains but eventually collapses into a planar GB with the probability
$(2/3)(1-P_{12})P_{c}$.

The bottom row on the diagram (Fig.~\ref{fig:Event-diagram}) summarizes
the probabilities of disappearance of the initial triangular bump
along the four different routes. Since their sum must be equal to
$P_{c}$, we have the equation
\[
P_{c}=\dfrac{1}{3}P_{10}+\dfrac{1}{3}(1-P_{10})P_{c}+\dfrac{2}{3}P_{12}P_{r}P_{c}+\dfrac{2}{3}(1-P_{12})P_{c},
\]
which is solved for $P_{c}$: 
\begin{equation}
P_{c}=\dfrac{P_{10}}{P_{10}+2P_{12}(1-P_{r})}.\label{eq:collapse-prob-1}
\end{equation}

The remaining unknown is the return probability $P_{r}$. The problem
of finding $P_{r}$ can be formulated as follows. A kink pair attempts
to grow starting from a two-node state. The growth can be described
as a driven random walk of the number of nodes in the pair starting
from two. At each step, the number of nodes can increase by one with
the probability $P_{12}$, decrease by one with the probability $P_{21}<P_{12}$,
or not change. We must find the probability that the kink pair eventually
collapses into a single-node state (triangular bump). This problem
is equivalent to the cliff-hanger problem of a man making random steps
starting one step away from a cliff \citep{Mosteller-1965}. The probabilities
of steps away and towards the cliff are $p>1/2$ and $(1-p)<1/2$,
respectively. The known solution of this problem is that the probability
of falling off the cliff is $(1-p)/p$ \citep{Mosteller-1965}. This
solution maps onto our kink pair problem by identifying $p=P_{12}/(P_{12}+P_{21})$.
It follows that
\begin{equation}
P_{r}=\dfrac{P_{21}}{P_{12}}.\label{eq:return prob}
\end{equation}
Inserting this solution into Eq.(\ref{eq:collapse-prob-1}), we obtain
\begin{equation}
P_{c}=\dfrac{P_{10}}{P_{10}+2(P_{12}-P_{21})},\label{eq:collapse-prob-1-1}
\end{equation}
and thus
\begin{equation}
P_{s}=1-P_{r}=\dfrac{2(P_{12}-P_{21})}{P_{10}+2(P_{12}-P_{21})},\label{eq:Escape_prob-1}
\end{equation}
which is Eq.(\ref{eq:Escape_prob}) of the main text.

Equation (\ref{eq:collapse-prob-1-1}) was verified by independent
KMC simulations implementing the process presented in Fig.~\ref{fig:Event-diagram}.


\newpage\clearpage{}

\begin{table}
\begin{tabular}{|l|c|c|}
\hline 
Variable & Physical & Normalized\tabularnewline
\hline 
\hline 
Coordinates & $x$, $z$ & $\xi=x/a$,\quad{}$\varsigma=z/a$\tabularnewline
\hline 
Time & $t$ & $\tau=t\nu_{0}$\tabularnewline
\hline 
Velocity & $v$ & $\eta=v/a\nu_{0}$\tabularnewline
\hline 
Temperature & $T$ & $\theta=k_{B}T/E_{0}$\tabularnewline
\hline 
Driving force & $F$ & $\varphi=Fa/E_{0}$\tabularnewline
\hline 
GB energy & $\gamma$ & $\sigma=\gamma a^{2}/E_{0}$\tabularnewline
\hline 
GB stiffness & $\Lambda$ & $\lambda=\Lambda a^{2}/E_{0}$\tabularnewline
\hline 
Mean squared GB width & $\overline{w^{2}}$ & $\overline{\omega^{2}}=\overline{w^{2}}/a^{2}$\tabularnewline
\hline 
\end{tabular}

\caption{Physical and normalized (dimensionless) variables in the 2D GB model.\label{tab:Reduced-variables}}

\end{table}

\newpage\clearpage{}

\newpage\clearpage{}

\newpage\clearpage{}

\end{document}